\documentclass[11pt]{article}
\usepackage[utf8]{inputenc}
\usepackage[T1]{fontenc}
\usepackage{mathptmx}
\usepackage{textcomp}
\usepackage{amsmath}
\usepackage{graphicx}
\usepackage{longtable}
\usepackage{pdflscape}
\usepackage{setspace}
\usepackage{caption}
\usepackage[letterpaper,margin=1in]{geometry}
\usepackage{microtype}
\usepackage[hidelinks]{hyperref}
\usepackage{titlesec}
\titleformat{\section}{\large\bfseries\raggedright}{\thesection}{0.6em}{}
\titleformat{\subsection}{\normalsize\bfseries\raggedright}{\thesubsection}{0.6em}{}
\titleformat{\subsubsection}{\normalsize\bfseries\itshape\raggedright}{\thesubsubsection}{0.6em}{}
\titlespacing*{\section}{0pt}{1.4ex plus .3ex}{0.6ex}
\titlespacing*{\subsection}{0pt}{1.1ex plus .3ex}{0.4ex}
\begin{document}

\begin{center}
{\Large\bfseries Nanoscale Storage of Incompatible Elements at Olivine Grain Boundaries in Natural Basalts}\\[1.1em]
{\large Wenhao Zhao$^{1}$, Reid Cooper$^{1}$, Stephen Parman$^{1}$, Austin Akey$^{2}$, Greg Hirth$^{1}$}\\[0.7em]
$^{1}$Department of Earth, Environmental and Planetary Sciences, Brown University, Providence, RI 02912, USA\\
$^{2}$Center for Nanoscale Systems, Harvard University, Cambridge, MA 02138, USA\\[0.2em]
Correspondence: wenhao\_zhao@brown.edu
\end{center}
\vspace{0.8em}

\section*{Abstract}
Grain boundaries are pervasive in polycrystalline olivine, yet their structure and trace-element chemistry under natural magmatic conditions remain poorly constrained. We characterize boundaries in undeformed olivine aggregates from Piton de la Fournaise (La Réunion) and Mauna Loa (Hawaii) by correlating electron backscatter diffraction, transmission electron microscopy, and atom probe tomography. The boundaries are crystalline, with a structural width of only \textasciitilde{}1 nm and no continuous glassy film resolved. Ca, Al, P, Na, and Ti are selectively enriched, whereas the well-fitted major cations Mg, Fe, Ni, and Mn remain comparatively homogeneous. Several boundary-enriched mass windows coincide with nominal REE molecular-ion positions, but unresolved isobars and the absence of a diagnostic isotope envelope preclude secure elemental assignments. The divalent cations follow an ionic-size-misfit trend, while aliovalent departures require additional, presently unconstrained contributions.

The Ca interfacial excess is 2.94--2.97 atoms\,nm\textsuperscript{--2}, equivalent to 0.327--0.330 monolayers, and plots near the upper published 1523 K equilibrium two-dimensional segregation isotherm of Hiraga et al. (2004), within the range of basaltic olivine crystallization temperatures. This agreement supports boundary formation during crystal growth, although synneusis followed by sufficient high-temperature residence could produce a similar signature. Ca has cumulative-fraction chemical widths of \textasciitilde{}6--9 nm rather than being confined to a saturated single plane, and the chemical widths of the most strongly enriched elements reach \textasciitilde{}10--15 nm, demonstrating decoupling between the narrow structural core and broader chemical segregation. Relative to published atom-probe data on deformed olivine, these undeformed boundaries preserve broader halos, although cross-study analytical differences make that contrast suggestive rather than definitive. Olivine grain boundaries therefore constitute a distinct nanoscale reservoir that should be included alongside crystal interiors and residual melt in trace-element mass balances of olivine-rich aggregates.

\section*{Highlights}
\begin{itemize}\setlength{\itemsep}{0pt}
\item Correlated TEM--APT resolves grain boundaries in undeformed natural basaltic olivine
\item Ca, Al, P, Na and Ti are enriched at olivine grain boundaries
\item Divalent segregation follows size misfit; aliovalent terms remain unconstrained
\item A \textasciitilde{}1 nm crystalline core is decoupled from a wider (\textasciitilde{}10 nm) chemical halo
\item Ca coverage matches equilibrium segregation near the crystallization temperature
\end{itemize}

\vspace{0.5em}
\noindent\textbf{Keywords:} olivine; grain boundaries; atom probe tomography; trace-element segregation; incompatible elements

\section{Introduction}

The differing geochemical properties and affinities that trace elements exhibit among coexisting reservoirs are the foundation of trace-element geochemistry (Goldschmidt, 1937). In practice, these properties manifest as a measurable distribution and abundance of trace elements among phases, so characterizing where and how trace elements reside in real systems remains a continuing and fundamental task of geochemistry. Most commonly, this partitioning behavior is described between two phases --- for example, by the Nernst partition coefficient between a crystal and its melt (McIntire, 1963; Blundy \& Wood, 1994, 2003), or between two crystals (Onuma et al., 1968).

In nature, however, minerals most commonly occur as polycrystalline aggregates whose constituent grains are separated by internal interfaces: grain boundaries between crystals of the same mineral, and phase boundaries between different minerals. Both grain boundaries (GBs) and phase boundaries are an intrinsic component of essentially every rock (Marquardt et al., 2015), and, because their atomic configuration departs from that of the adjacent crystal interiors (Cantwell et al., 2014), they also have the potential to act as reservoirs for trace elements that the lattice excludes (Hiraga et al., 2003, 2004). Understanding whether, and how, the lattice's excluded elements accumulate at these interfaces is the central motivation of the present study.

That chemistry confined to interfaces only a few nanometers thick should matter at the macroscopic scale --- particularly in natural magmatic growth settings --- is not self-evident; three considerations, each anchored in prior work, motivate our focus on trace-element behavior at these boundaries. First, although an individual boundary is only nanometers thick, boundaries are ubiquitous, and their cumulative effect can be disproportionate: they influence the bulk physical properties of an aggregate (Hirth \& Kohlstedt, 2003; Marquardt et al., 2015) and, in mass-balance terms, can constitute a chemical reservoir that --- though volumetrically minor and dilute --- carries geological significance out of proportion to its size (Hiraga et al., 2004), much as trace reservoirs such as the lunar urKREEP (the final residual liquid of the lunar magma ocean, enriched in K, rare-earth elements (REE) and P; Warren \& Wasson, 1979; Warren, 1985) or ore-forming concentrations do. Second, boundaries are not passive storage sites but obligatory pathways for diffusion, grain-boundary sliding and melt migration (Dohmen \& Milke, 2010; Hirth \& Kohlstedt, 2003; Waff \& Bulau, 1979), so their chemistry can modulate how an aggregate deforms, transports mass and equilibrates. Third, interfacial enrichment has until recently been understood largely through interfacial-thermodynamic theory (McLean, 1957; Lejček, 2010), with direct observation lagging behind because chemical information at the nanometer scale was difficult to recover; the application of atom probe tomography (APT) to geological and planetary materials (Reddy et al., 2020; Saxey et al., 2018) has now eased much of that limitation, so that boundaries once averaged over can be examined directly.

Among the many kinds of interface, we focus here on a single subdivision --- grain boundaries in olivine from basaltic magmas --- a case that is both important and representative in two respects. From the mineral side, olivine is among the earliest phases to crystallize from basalt and the dominant mineral of Earth's upper mantle (Ringwood, 1975); its aggregates are abundant, compositionally well characterized and, being single-phase, allow grain-boundary storage to be isolated from the complications of multi-mineral assemblages. From the interface side, olivine's strong rejection of incompatible elements makes its boundaries an especially sensitive place to look for interfacial storage, while GBs more broadly govern diffusion, deformation and melt connectivity (Hirth \& Kohlstedt, 2003; von Bargen \& Waff, 1986), so their chemistry carries consequences beyond the trace-element budget. We stress that this is only a starting point: in principle every mineral assemblage and every interface type --- grain boundaries and phase boundaries alike --- carries its own distinct interfacial chemistry, and olivine in basalt serves here as a tractable first case rather than a special one.

What has actually been resolved at the scale of an individual olivine grain boundary, by contrast, remains sparse and is dominated by deformed material. The most direct nanoscale chemical evidence comes from two atom-probe studies, both of \emph{deformed} olivine: Cukjati et al. (2019) measured grain- and phase-boundary chemical widths of $\sim$3.1--6.6~nm in experimentally deformed wehrlite, and Tacchetto et al. (2021) found segregation of Ca, Al, Ti, P, Mn, Fe, Na and Co to low-angle subgrain boundaries in naturally deformed olivine, increasing systematically with disorientation angle. Because both target deformed olivine, their boundary chemistry may already have been reset by dynamic recrystallization and dislocation-related processes, and need not record an equilibrium state. The complementary case --- chemistry inherited from growth at near-magmatic conditions in undeformed natural olivine --- is precisely the one that would distinguish whether interfacial storage is an intrinsic, rock-forming property of olivine aggregates or merely a by-product of deformation, yet it has not been examined at this scale. The stakes are not small: even though a boundary occupies a tiny volume --- for a grain size $d$ and an effective width $\delta$, the GB volume fraction is roughly $3\delta/d$, of order $10^{-3}$ --- enrichment factors of $10^{2}$--$10^{3}$ for elements with low lattice solubility can still make its contribution to the bulk inventory non-negligible (cf. Hiraga et al., 2004). Whether such storage operates in natural, undeformed basaltic olivine that preserves near-magmatic growth, how it is structurally hosted, and how large the stored fraction is, are the questions this study addresses.

As anticipated above, APT is especially well suited to this task. It yields three-dimensional reconstructions in which individual ions are positioned at sub-nanometer to nanometer-scale resolution and identified by time-of-flight mass spectrometry, so that the composition of a feature only a few nanometers across --- such as a grain boundary --- can be measured directly rather than averaged into the surrounding crystal or melt (Reddy et al., 2020). Under favorable, interference-free conditions APT can attain parts-per-million sensitivity, potentially reaching trace elements that are below the practical detection limits of conventional TEM-based energy-dispersive X-ray or electron energy-loss spectroscopy (Reddy et al., 2020; Saxey et al., 2018). In silicates, however, practical detectability and quantification are element- and ion-species-dependent: field evaporation, molecular-ion formation, background, peak overlaps, ranging choices, and reconstruction artifacts can all bias the recovered signal. APT does not require conventional external standards, but it should not be treated as uniformly sensitive across the periodic table, and light species such as hydrogen are particularly susceptible to residual-gas and preparation-related backgrounds (Valley et al., 2014). In insulating minerals the practical spatial resolution is commonly closer to one nanometer than to the sub-nanometer values reported for metals, but this remains sufficient to resolve the chemical structure of a single olivine grain boundary in three dimensions. A further challenge is that natural silicates produce mass-spectral overlaps --- in particular among Fe, Si, and their molecular ions --- which complicate accurate identification and quantification of trace species and require explicit treatment in this study.

In this study we combine TEM and APT to characterize GBs in olivine aggregates from two basaltic systems: Piton de la Fournaise (La R\'eunion) and Mauna Loa (Hawaii). TEM resolves the crystallographic structure of the boundary, while APT provides three-dimensional, sub-nanometer compositional maps with single-atom sensitivity; together they bridge the structural and chemical scales at which GBs operate. Both sample suites are undeformed and preserve near-magmatic olivine that grew in contact with its host melt. Our aims are to determine whether incompatible elements are systematically enriched at olivine GBs, to establish whether that enrichment resides in the crystalline boundary rather than in trapped melt, and to assess the consequences for trace-element mass balance.

\section{Samples and analytical methods}

\subsection{Piton de la Fournaise volcano, La Réunion}
The first set of olivine polycrystals analyzed in this study were obtained from Piton de la Fournaise, La Réunion, through Benoît Welsch. These clots exhibit a dendritic-growth texture, representing an endmember described by Welsch et al. (2013). For an in-depth discussion of the geological context, we refer readers to Michon et al. (2015), while providing only a brief overview here.

La Réunion Island consists of two primary volcanic edifices: the extinct Piton des Neiges (Salvany et al., 2012) and the active Piton de la Fournaise (Gillot \& Nativel, 1989). Both volcanoes produce aphyric basalts as well as olivine-rich porphyritic basalts. The latter, from which our samples originate, exhibit a broad compositional spectrum, encompassing magnesian basalts (9--12 wt\% MgO), picrites (>12 wt\% MgO), and oceanites (>20 wt\% MgO). Famin et al. (2009) proposed that olivine-rich basalts contain a mix of phenocrysts and xenocrysts, suggesting that their formation is influenced by complex magmatic processes, including magma mixing, crystal recycling, and ascent dynamics. These processes likely played a role in the development of the olivine polycrystal clots analyzed in this study.

The Réunion olivine analyzed in this study is shown in Figure 1(a--c). Figure 1(a) shows the polished olivine clot, composed of two euhedral crystals with a distinct separation at the center; EBSD (Results) classifies this interface as a high-angle grain boundary, but the archived Réunion export does not retain the rotation-axis metadata required for an independent full twin-law test. Figure 1(b) resolves the grain-boundary network under reflected light, with the prominent analyzed GB enclosed in a green open rectangle; this boundary appears long and planar, suggesting minimal deformation or curvature. Figure 1(c) presents a high-magnification view of this grain boundary, which appears as a sharp, continuous line with subtle variations in transparency and reflectivity along its trace; its nanoscale structure and chemistry are resolved by the TEM and APT analyses below. Samples for those analyses were extracted from this region, at or across the grain boundary.

\subsection{Mauna Loa, Hawaii}
Our second olivine samples came from Mauna Loa, Hawaii, and consist of polycrystalline clots whose juxtaposed grains could reflect co-growth or aggregation during magma storage and transport. We obtained these samples from Fred Anderson. More details on the local geology and olivine morphology can be found in Rhodes (1995) and Schwindinger \& Anderson (1989); here, we provide a brief introduction.

Mauna Loa, the largest active shield volcano on Earth, spans more than half of Hawaii's Big Island. Its structure is predominantly built from tholeiitic basalt, with frequent eruptions generating extensive lava flows that shape its broad, gently sloping topography. Olivine is a widespread mineral in Mauna Loa basalts, appearing both as individual crystals and as glomeroporphyritic aggregates. Such olivine-rich clots have commonly been interpreted as products of synneusis and crystal accumulation within magma reservoirs before eruption, although co-growth of adjacent crystals provides an alternative origin relevant to the present samples. Research indicates that olivine in Mauna Loa basalts crystallizes at varying depths, reflecting a complex magmatic plumbing system with multiple storage zones beneath the volcano (Wieser et al., 2025). The composition and three-dimensional morphology of olivine vary both within individual aggregates and across different lapilli. Although this variability has been used to infer aggregation of independently nucleated crystals, most crystals are attached along crystallographic faces, with their c-axes aligned either parallel or perpendicular, a geometry also compatible with growth that minimizes structural mismatch and interfacial energy (Schwindinger \& Anderson, 1989). We therefore treat synneusis and co-growth as alternative origins; distinguishing between them is not required for the grain-boundary analysis that follows.

Representative images of the Mauna Loa olivine clots examined in this study are shown in Figure 1(d--f). Figure 1(d) shows the clot, which consists of three euhedral olivine crystals, reflecting a largely unaltered crystallization history. Thermal cracks extend across the crystals, likely from cooling-induced fracturing, and the crystals host abundant composite inclusions---glass--chromite assemblages, gas bubbles, and discrete chromite. The analyzed grain boundary is outlined with a green rectangle under reflected light in Figure 1(e). Near this boundary, Figure 1(f) shows residual melt with nucleating chromite, recording late-stage melt interaction within the aggregate. Samples for TEM and APT were extracted at the location marked in Figure 1(f), at or across the grain boundary.

\begin{figure}[htbp]
\centering
\includegraphics[width=\textwidth]{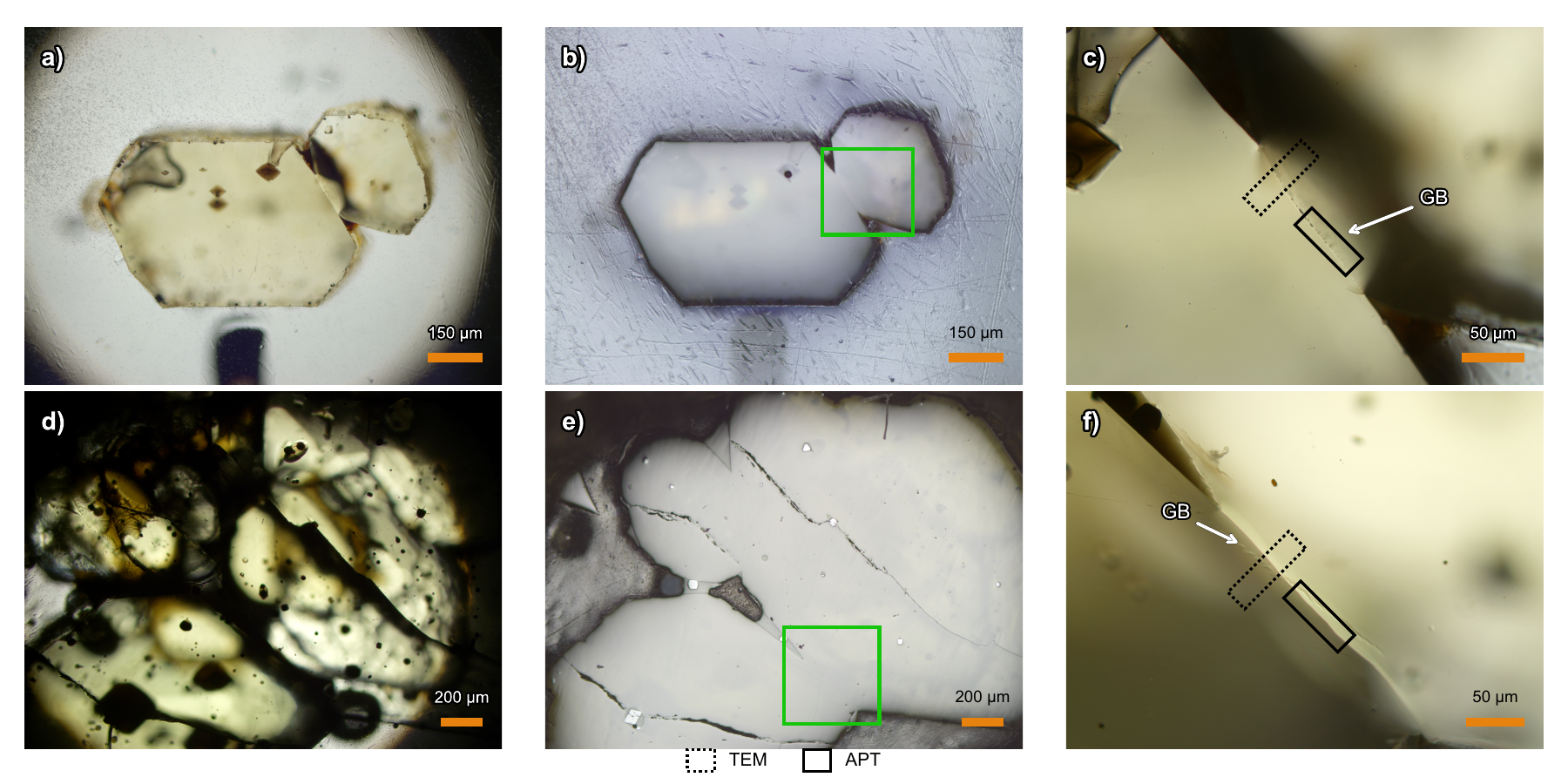}
\caption*{\textbf{Figure 1.} Optical petrography of the two analyzed olivine aggregates; top row R\'eunion (Piton de la Fournaise), bottom row Mauna Loa (Hawaii). (a, d) Plane-polarized overview of the polished olivine clot. (b, e) Reflected-light image resolving the grain-boundary network, with the green rectangle marking the region of interest that contains the analyzed boundary. (c, f) Close-up of the analyzed grain boundary; the TEM (dotted) and APT (open rectangle) lift-out locations and the boundary (GB, arrow) are marked.}
\end{figure}

\subsection{EBSD and FIB-TEM-APT}

\subsubsection{EBSD}
Electron backscatter diffraction (EBSD) was used to determine crystallographic orientations and misorientation relationships across the selected boundaries. Data were acquired using a Zeiss Gemini 500 field-emission scanning electron microscope equipped with an Oxford Symmetry S2 detector at the Center for Nanoscale Systems (CNS), Harvard University. Conventional reflection EBSD was performed on the planar surfaces of FIB-prepared lamellae mounted for subsequent TEM work, rather than through an electron-transparent TEM foil. Maps were collected at 20 kV, a 15 mm working distance, and the 0.125 $\mu$m step recorded in the map metadata (Figure S1); this raster samples the two grain orientations but does not imply 125 nm spatial resolution at the boundary. Oxford Instruments AZtec and MTEX were used to generate orientation maps, calculate disorientation operators, and classify boundary segments. The reported angle and $\pm$ value are the mean and one standard deviation, respectively, of the mapped boundary-segment disorientation values, not the analytical uncertainty of a single orientation measurement. Where the full operator was retained (Mauna Loa), the rotation angle and axis were compared jointly with the olivine twin operators tabulated by Azevedo and Nespolo (2017).

\subsubsection{FIB-TEM-APT}
Grain boundaries of interest were selected for Focused Ion Beam (FIB) preparation. APT and TEM samples were extracted using a dual-beam FEI Helios 660 FIB-SEM, equipped with an Omniprobe AutoProbe 200 micromanipulator, at Harvard CNS. To ensure precise grain boundary positioning, boundaries were marked with carbon or electron beam-deposited platinum before shaping TEM foils and APT needles.

To minimize Ga\textsuperscript{+} ion beam damage, a thin platinum (Pt) protective layer was deposited over the region of interest before milling. For TEM sample preparation, lamellae were extracted from the same grain boundaries selected for APT, allowing a direct correlation between structural and compositional analyses. TEM foils were thinned to \textasciitilde{}100 nm using a 30 kV Ga-ion beam, followed by polishing at 5 kV and 2 kV to reduce surface damage and enhance specimen quality.

A standard APT FIB lift-out procedure was applied (Thompson et al., 2007) with final milling performed at 30 kV, gradually reducing the beam current from 2.9 nA to 16 pA to achieve precise needle shaping. APT tips were sculpted into a long, narrow conical geometry atop a tapered cylindrical column to stabilize field evaporation. Each lift-out contained sufficient material to produce 4--6 needle-shaped specimens for APT analysis.

TEM characterization was conducted using a JEOL ARM 200F scanning transmission electron microscope (STEM) at Harvard CNS, focusing on crystal defects, grain boundary structures, and nanoscale phase segregation. Bright-field TEM, dark-field TEM, and high-angle annular dark-field imaging were used to investigate crystallinity, dislocation density, and interfacial textures.

APT analysis was performed using a CAMECA LEAP 4000X HR at Harvard CNS to obtain high-resolution compositional data from selected grain boundaries. Data were collected in laser-pulsed mode at a pulse frequency of 100 kHz and a specimen base temperature of $\sim$44 K. The laser pulse energy and target detection rate were set per specimen to optimize field evaporation: 50 pJ and 0.5\% (0.005 ions per pulse) for the R\'eunion needle, and 60 pJ and 1.0\% (0.010 ions per pulse) for the Mauna Loa needle. For each locality (R\'eunion and Mauna Loa), a single needle-shaped specimen containing the analyzed grain boundary was run to completion --- the reconstructions shown in Figures 3 and S5 --- so the dataset comprises one fully analyzed boundary per locality. Each acquisition comprised of order $10^{7}$--$10^{8}$ ions ($\sim$208 million for the R\'eunion reconstruction and $\sim$22.5 million for Mauna Loa) collected over a reconstructed length of \textasciitilde{}1.16 $\mu$m (R\'eunion) and \textasciitilde{}250 nm (Mauna Loa). The R\'eunion reconstruction is the more complete dataset --- an order of magnitude more ions over a $\sim$5$\times$ longer boundary --- and is therefore the primary dataset shown in the main text (Figures 3--5) and used for the rare-earth analysis; the shorter Mauna Loa reconstruction is presented as a confirmatory case in the Supplement (Figures S5, S7). The mass resolving power ($M/\Delta M$ at full width at half maximum, FWHM) was \textasciitilde{}1200 at the \textsuperscript{24}Mg\textsuperscript{2+} peak, and the detection efficiency of the LEAP 4000X HR reflectron is \textasciitilde{}36\%. We note that the Mauna Loa (Hawaii) specimen was destroyed after repeated APT and TEM specimen preparation, so no further measurements could be made on the same material. APT datasets were reconstructed and analyzed using AP Suite 6 software (CAMECA). The three-dimensional reconstruction geometry was calibrated against SEM images of each specimen needle acquired during preparation; no independent crystallographic plane-spacing calibration was applied.

For the crystallographic overlays in Figure S10, detector-hit-density (desorption) maps were generated separately from subvolumes on each side of each boundary. Pole centers and associated zone traces were selected on the density maps and manually indexed against the orthorhombic olivine stereographic geometry. An index was retained only where one orientation simultaneously reproduced several non-collinear poles and their intervening zone-line angular relationships; isolated line matches were not accepted. The red lines in Figure S10 are therefore calculated zone traces for the accepted orientation, not direct images of atomic planes. No spatial-distribution-map or Fourier-transform measurement was used for this indexing, and the overlays are used only as a crystallographic consistency check rather than as an independent reconstruction calibration.

Because the two specimens were analyzed under different laser energies and target detection rates, and because the Réunion reconstruction contains roughly an order of magnitude more ions than the Mauna Loa reconstruction, small differences between localities may include acquisition- and reconstruction-dependent contributions. We therefore use the second boundary as a qualitative replication of the principal segregation pattern rather than treating inter-locality differences in peak height or width as precisely resolved.

\subsection{Boundary thickness measurements}
The Gibbsian interfacial-excess calculation follows the atom-probe formulation of Krakauer and Seidman (1993).

There are two primary methods for measuring chemical widths based on APT data: the "one-dimensional (1-D) compositional profile" method and the "proxigram" method. For a more detailed discussion, readers are referred to Hellman et al. (2000) and Gault et al. (2012).

In this study we used the 1-D compositional profile method. The proxigram method relies on iso-concentration surfaces that are rarely perfectly planar; their small-scale topographic irregularities (bumps, convexities, concavities) introduce significant uncertainty into the statistics computed around them. The 1-D method has its own limitation---the region of interest (ROI) is selected manually and may not align exactly with the boundary---but we judge this minor relative to the distortions of the proxigram approach. Adopting the 1-D method throughout also keeps our measurements consistent with previously published data, allowing meaningful comparison of relative grain-boundary thicknesses.

For the 1-D compositional profile method, we follow the approach outlined by Cukjati et al. (2019). We define a subvolume within the (x, y, z) chemical dataset with an orthorhombic geometry that encompasses the grain or phase boundary of interest. One of the principal axes of the orthorhombic subvolume---or, in some cases, the cylindrical axis---is oriented perpendicular to the boundary to ensure accurate profiling. Compositional variations along the boundary normal are quantified by averaging chemical concentrations within evenly spaced bins perpendicular to the normal direction. In this study, chemical profiles were calculated using a bin spacing of 0.10 nm, while data within 1 nm of the exterior surface were excluded to minimize artifacts from edge effects.

The active R\'eunion profile used in Figures 5, 7 and 8 is the AP Suite export ``R13\_05409-v01 -- Cube ID 6 -- 1D Concentration Profile -- Y-axis.'' It is the Y-axis profile normal to the boundary along the green arrow in Figure 4, spans 63.4 nm at 0.10 nm spacing, and replaces the obsolete Cube ID 4 worksheet used in an earlier draft. The export does not retain the lateral dimensions of Cube ID 6; its footprint and extraction direction are therefore documented by the reconstruction view in Figure 4 rather than inferred. The Mauna Loa profile is ``R13\_05578-v01 -- Cube ID 1 -- 1D Concentration Profile -- X-axis,'' spans 49.8 nm at 0.10 nm spacing, and follows the green marker in Figure S6b. All references to Fe in these profiles denote the total elemental Fe column reported by AP Suite, not a single Fe ion or molecular-ion channel.

For the Ca interfacial-excess calculation, the Ca occupancy of olivine M sites is defined as $X_{\mathrm{Ca}}=\mathrm{Ca}/(\mathrm{Mg}+\mathrm{Fe}+\mathrm{Ca}+\mathrm{Mn}+\mathrm{Ni}+\mathrm{Co})$ using the elemental atom-percent columns. The boundary centre is the maximum $X_{\mathrm{Ca}}$ value. The grain-interior baseline $X_{GM}^{\mathrm{Ca}}$ is the mean of bins at $|d|\geq8$ nm from that centre, and the signed difference $X_{\mathrm{Ca}}(d)-X_{GM}^{\mathrm{Ca}}$ is integrated across the full exported profile. Multiplication by the olivine M-site number density, $N_M=27.9$ nm$^{-3}$, gives $\Gamma_{\mathrm{Ca}}$ in atoms\,nm$^{-2}$; division by the equivalent M-site areal density, $N_M^{2/3}=8.98$ nm$^{-2}$, gives the equivalent monolayer coverage plotted in Figure 7. Counting uncertainty is propagated from the AP Suite 1$\sigma$ columns by Monte Carlo resampling; analytical sensitivity is evaluated by varying the matrix exclusion to 6, 8 and 10 nm and trimming 0, 1 and 2 nm from both profile ends. These components are combined in quadrature. The calculation, intermediate values and plotted uncertainties are supplied in \texttt{data/Ca\_interfacial\_excess.csv} and generated by \texttt{code/make\_figure7.py}; unquantified systematic APT compositional bias is not included in the error bars.

Figure 4 illustrates the approach using Ca concentration across the Réunion olivine grain boundary as an example. A cubic region of interest was selected perpendicular to the GB (Figure 4a), along which a 1-D compositional profile was extracted. The right panel (Figure 4b) displays the measured Ca concentration (black curve) as a function of distance, with the GB region highlighted in gray. To define the GB chemical thickness, we used a normalized cumulative fraction analysis of Ca distribution (red curve). Three linear fits were applied to the fraction curve, corresponding to the two olivine grain interiors (y1 and y3) and the GB enrichment zone (y2). The intersections of the fitted lines define the boundaries of the GB, with the distance between these intersections representing the GB chemical thickness. This method provides a robust and reproducible measure of GB thickness, minimizing subjective bias in selecting boundary positions. The approach is applicable to other trace elements and enables a quantitative comparison of GB segregation behavior across different samples.

\section{Results}

\subsection{EBSD}
Figure S1 presents EBSD analyses of olivine samples from both Réunion and Hawaii, highlighting grain boundary characteristics and crystallographic orientations.

In Figure S1(a, b), the EBSD maps combine band contrast (BC), inverse pole figure (IPF) coloring, and GB mapping. The band contrast map depicts variations in diffraction quality, where lower-contrast regions indicate potential grain boundaries or surface imperfections. The IPF coloring, referenced to the Z-axis, reveals a largely uniform orientation in the primary grain, suggesting minimal internal misorientation.

Figure S1(c, d) displays the disorientation-angle distributions, which peak at high angles. The mapped boundary-segment values are 50.09\textdegree{} $\pm$ 0.65\textdegree{} (mean $\pm1\sigma$) for R\'eunion and 51.68\textdegree{} $\pm$ 0.14\textdegree{} for Mauna Loa. The retained Mauna Loa operators span 51.52--51.76\textdegree{} about an axis within 2.62--3.17\textdegree{} of $[\bar{1}00]$; this angle--axis combination does not match the established olivine twin operators (Azevedo \& Nespolo, 2017). The archived R\'eunion export contains the angle distribution but not the corresponding rotation-axis/plane metadata, so we classify that interface as a high-angle boundary but do not claim an independent full twin-law exclusion for it.

\subsection{TEM}
Based on the TEM analysis presented in Figure 2, the microstructural characteristics of the Réunion olivine grain boundary were examined in detail. Figure 2(a) shows a low-magnification bright-field image of the boundary separating two olivine grains, and Figure 2(b) a higher-magnification view. The high-resolution TEM (HR-TEM) image (Figure 2c) provides an atomic-scale view of the boundary, with an inset selected-area electron diffraction (SAED) pattern confirming the crystallographic orientation of the adjacent grain. Intensity-profile analysis (Figure 2d) indicates an apparent GB thickness of approximately 1.15 nm, and lattice images of the two adjacent grains (Figure 2e,f) show well-ordered crystalline frameworks extending to the boundary on both sides.

The Hawaiian olivine shows a comparable GB structure (Figure S2). The HR-TEM image, optimized for thickness measurement, reveals a GB thickness of \textasciitilde{}1.05 nm, and the corresponding SAED pattern confirms the crystallinity of the adjacent grains; the intensity profile across the boundary (Figure S2d) gives the same measure.

High-resolution STEM images of both boundaries (Figure S3) resolve periodic lattice contrast up to the interface on both sides, with no continuous amorphous or glassy film resolved within the spatial and projection limits of the images. Thus, despite the apparent nanoscale thickness of the GB (\textasciitilde{}1.05--1.15 nm), the cooled boundaries are predominantly crystalline. Indexed poles and calculated zone traces on APT detector-hit-density maps from both sides of the boundaries (Figure S10) provide a crystallographic consistency check, although they are not direct images of lattice planes.

\begin{figure}[htbp]
\centering
\includegraphics[width=\textwidth]{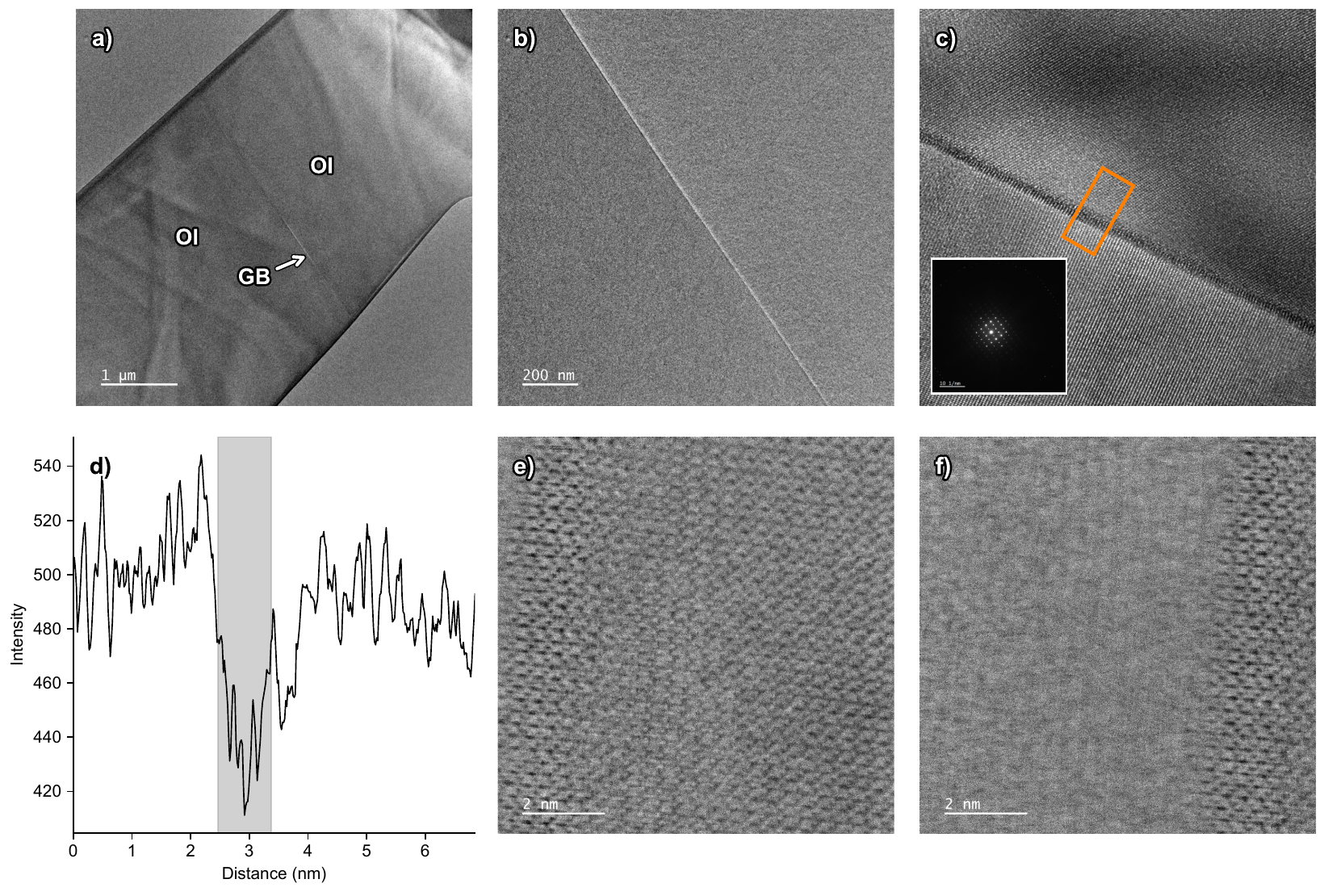}
\caption*{\textbf{Figure 2.} TEM of the R\'eunion olivine grain boundary. (a) Low-magnification bright-field image showing the boundary (GB) separating two olivine grains (Ol). (b) Higher-magnification bright-field view of the boundary. (c) High-resolution TEM (HR-TEM) image across the boundary; inset: selected-area electron diffraction (SAED) pattern of the adjacent grain, and the orange box marks the analyzed region. (d) Intensity profile across the boundary (long edge perpendicular to the GB; boundary region shaded), giving an apparent structural GB thickness of \textasciitilde{}1.15 nm. (e, f) HR-TEM lattice images showing well-ordered crystalline fringes extending to the boundary from the two adjacent grains. The corresponding TEM data for the Mauna Loa (Hawaii) boundary are shown in Figure S2, and high-resolution STEM images of both boundaries in Figure S3.}
\end{figure}

\subsection{APT}

\subsubsection{Ion species and bulk compositions}
The identification of ion species based on mass spectrum peaks is presented in Table S1 and Figure S4. As expected, the primary elements detected in olivine are Mg, Fe, Si, and O, with minor and trace elements including Ca, Al, Ni, Mn, and P. A summary of the bulk compositions of Réunion and Hawaii olivine, along with analytical uncertainties, is provided in Table 1.

\begin{table}[htbp]
\centering
\caption*{\textbf{Table 1.} APT-measured bulk compositions of the Réunion and Hawaiian olivine specimens (atomic \%), ordered by decreasing Réunion abundance. Values are obtained from the ranged atom-probe reconstructions (Table S1); uncertainties are 1$\sigma$ counting statistics. The compositions are semi-quantitative: absolute values are affected by oxygen loss and Fe--Si isobaric overlaps intrinsic to silicate APT (see text), so they are used here only to establish relative, cross-boundary behavior, not as calibrated abundances.}
\small
\begin{tabular}{lcc}
\hline
Ion type & Réunion (at.\%) & Hawaii (at.\%) \\
\hline
O & 46.48 $\pm$ 0.06 & 48.63 $\pm$ 0.20 \\
Mg & 31.55 $\pm$ 0.06 & 30.66 $\pm$ 0.18 \\
Si & 12.71 $\pm$ 0.04 & 13.71 $\pm$ 0.14 \\
Fe & 6.36 $\pm$ 0.03 & 5.94 $\pm$ 0.09 \\
H\textsuperscript{a} & 2.41 $\pm$ 0.02 & 0.645 $\pm$ 0.035 \\
Ca & 0.121 $\pm$ 0.004 & 0.152 $\pm$ 0.016 \\
Al & 0.089 $\pm$ 0.004 & 0.052 $\pm$ 0.012 \\
Ni & 0.088 $\pm$ 0.004 & 0.068 $\pm$ 0.012 \\
Mn & 0.068 $\pm$ 0.003 & 0.068 $\pm$ 0.011 \\
P & 0.020 $\pm$ 0.003 & 0.014 $\pm$ 0.008 \\
Ti & 0.018 $\pm$ 0.003 & 0.013 $\pm$ 0.008 \\
Na & 0.0074 $\pm$ 0.0014 & 0.0065 $\pm$ 0.0040 \\
Co & 0.014 $\pm$ 0.002 & 0.0097 $\pm$ 0.0068 \\
Cr & 0.0032 $\pm$ 0.0011 & 0.0028 $\pm$ 0.0033 \\
K & 0.0004 $\pm$ 0.0013 & 0.0006 $\pm$ 0.0036 \\
\hline
\end{tabular}

\vspace{0.3em}
{\footnotesize \raggedright
\textsuperscript{a}\,The apparent hydrogen is strongly affected by residual-gas and field-evaporation artifacts and is not a reliable structural abundance (cf.\ Figure 5).\\
Rare-earth elements are omitted because none is identified securely enough for elemental quantification. Several boundary-enriched candidate windows coincide with nominal REE molecular-ion positions, including 83.96 Da, but the diagnostic isotope envelopes are unresolved and alternative molecular-cluster carriers remain possible (Supplementary Text S1).\par}
\end{table}

Taking the Réunion olivine as an example (Table 1), the reported counting-statistics precision is approximately 0.12\% relative (1$\sigma$) for O and about 13\% relative for the lower-abundance P signal. These values do not include systematic uncertainties from oxygen loss, peak overlaps, ranging, field evaporation, or reconstruction. The measured O abundance (46.5 at.\% versus the ideal olivine value of 57.1 at.\%) and Si abundance (12.7 at.\% versus 14.3 at.\%) show substantial and moderate deficits, respectively, consistent with known compositional biases in silicate APT (Mancini et al., 2014; Reddy et al., 2020). Normalization after oxygen loss correspondingly inflates the apparent cation fractions, so the measured Mg\,+\,Fe total (37.9 at.\%) should not be compared directly with the ideal total-atom value of 28.6 at.\% as though it were a calibrated bulk composition. Because no independent electron-probe microanalysis (EPMA) compositions are available for these grains, the forsterite content is estimated directly from the APT Mg/(Mg\,+\,Fe) ratio, giving \textasciitilde{}Fo$_{83-84}$ (consistent with the \textasciitilde{}6 at.\% boundary-region Fe of Figure S7); we treat this as an approximate, APT-derived estimate rather than a calibrated composition, given the Fe--Si overlaps and oxygen-loss artifacts noted above.

The major peak overlap issues in olivine arise from similarities in the mass-to-charge ratios of Fe and Si ion species. For example, Fe\textsuperscript{+} (m/e = 55.93) overlaps with Si\textsubscript{2}\textsuperscript{+} (m/e = 55.00), and Fe\textsuperscript{2+} (m/e = 27.97) overlaps with Si\textsuperscript{+} (m/e = 28.09). These overlaps can complicate accurate elemental quantification of mafic minerals under APT and require careful peak deconvolution techniques to distinguish the contributions of each species. However, this issue is less critical in our study, as our primary focus is on the distribution of incompatible elements within GBs. The uncertainty in bulk composition does not significantly affect the interpretation of GB segregation, which rests on the relative enrichment across the boundary rather than on absolute bulk values.

\subsubsection{Grain boundary characterization}
The APT results presented in Figure 3 and Figure S5 provide a comparative analysis of the nanoscale chemical distributions in Réunion and Hawaiian olivine samples, respectively. Each map represents the spatial distribution of individual ion species, with one dot corresponding to a detected ion.

In Figure 3, which illustrates the Réunion olivine sample, the total ion distribution (upper-left panel) provides an overview of the sample composition; the maps show an $\sim$500 nm section around the grain boundary, cropped from the full $\sim$1.16 $\mu$m reconstruction. The upper-row maps of Mg\textsuperscript{+}, FeO\textsuperscript{+}, Si\textsuperscript{2+}, O\textsuperscript{+}, Cr\textsuperscript{+}, Ni\textsuperscript{+}, and Co\textsuperscript{+} indicate relatively uniform distributions without significant segregation. In contrast, the lower-row Ca\textsuperscript{2+}, AlO\textsuperscript{+}, OH\textsuperscript{+}, Na\textsuperscript{+}, PO\textsuperscript{+}, and TiO\textsuperscript{+} channels exhibit planar enrichment. The panel labelled SmO\textsuperscript{2+} is retained as a map of the nominal 83.96 Da candidate window; its spatial enrichment does not establish that Sm is the carrier.

Similarly, Figure S5 displays the APT analysis of the Hawaiian olivine sample, with a slightly shorter analyzed volume (\textasciitilde{}250 nm). The upper-row elements (Mg\textsuperscript{+}, FeO\textsuperscript{+}, Si\textsuperscript{2+}, O\textsuperscript{+}, Cr\textsuperscript{+}, Ni\textsuperscript{+}, and Co\textsuperscript{+}) again show a relatively homogeneous distribution. However, in the lower-row maps, Ca\textsuperscript{2+}, AlO\textsuperscript{+}, Na\textsuperscript{+}, and TiO\textsuperscript{+} display significant planar enrichment, aligning with the GBs, while OH\textsuperscript{+}, PO\textsuperscript{+}, and SmO\textsuperscript{2+} appear more homogeneously dispersed.

\begin{figure}[htbp]
\centering
\includegraphics[width=0.95\textwidth]{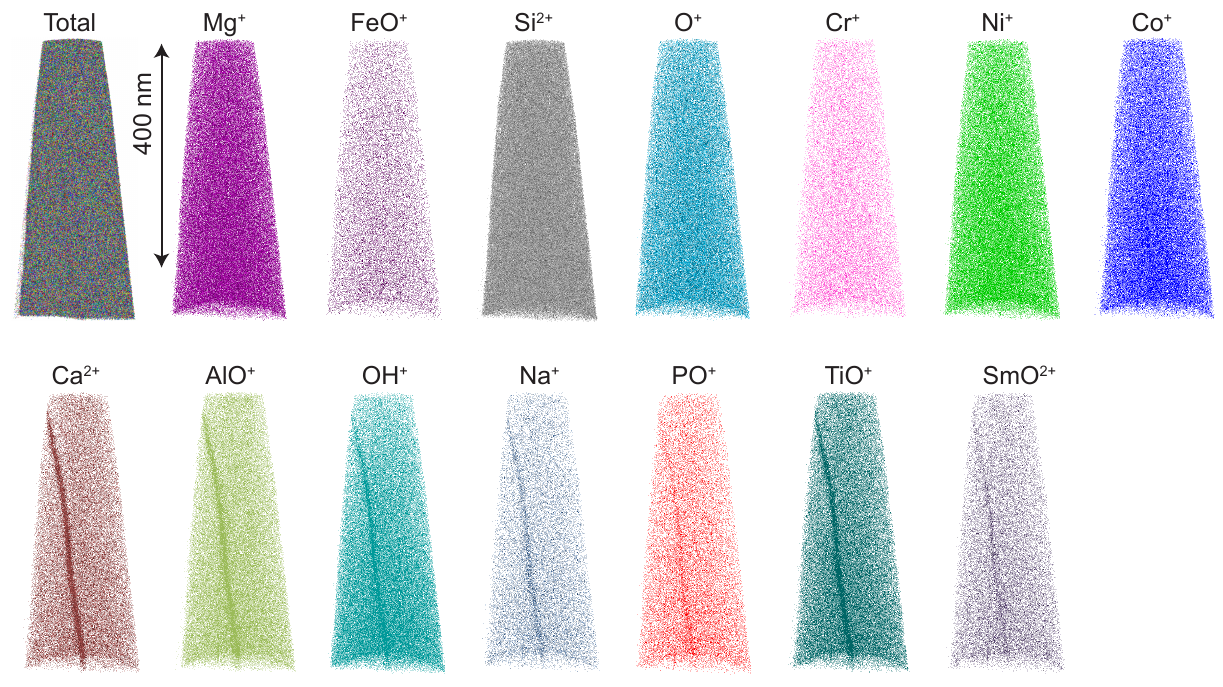}
\caption*{\textbf{Figure 3.} Atom probe tomography (APT) ion maps of the R\'eunion olivine needle (an $\sim$500 nm section around the grain boundary, cropped from the full $\sim$1.16 $\mu$m reconstruction), viewed face-on to the grain boundary so that the boundary plane is seen edge-on. The top-left panel shows all ranged species together (one color per ion type). The remaining top-row maps (Mg\textsuperscript{+}, FeO\textsuperscript{+}, Si\textsuperscript{2+}, O\textsuperscript{+}, Cr\textsuperscript{+}, Ni\textsuperscript{+}, Co\textsuperscript{+}) are homogeneous across the needle, whereas the confirmed minor-element channels Ca\textsuperscript{2+}, AlO\textsuperscript{+}, OH\textsuperscript{+}, Na\textsuperscript{+}, PO\textsuperscript{+} and TiO\textsuperscript{+} show planar enrichment. The panel labelled SmO\textsuperscript{2+} maps the nominal 83.96 Da candidate window. That window is boundary enriched, but its diagnostic \textsuperscript{147}SmO\textsuperscript{2+}/\textsuperscript{149}SmO\textsuperscript{2+}/\textsuperscript{152}SmO\textsuperscript{2+} isotope envelope is not resolved and alternative co-segregating molecular clusters cannot be excluded; it is therefore not treated as a confirmed Sm detection.}
\end{figure}

To first order, these APT datasets confirm the enrichment of Al, Ca, Na, and Ti at the GBs in both Réunion and Hawaiian olivine. Figure 4 and Figure S6 show Ca and Al iso-concentration surfaces at selected visualization thresholds (0.7 and 0.6 at.\% for Réunion; 0.8 and 0.7 at.\% for Mauna Loa); these thresholds delineate the interface and are not peak-concentration measurements. The quantitative 1-D profiles give Réunion point maxima of 0.63 at.\% Ca and 0.50 at.\% Al and Mauna Loa maxima of 1.00 at.\% Ca and 0.45 at.\% Al. Because individual-bin maxima are noise sensitive, comparisons below use the mean within $\pm2$ nm of the Ca-defined boundary centre.

\begin{figure}[htbp]
\centering
\includegraphics[width=\textwidth]{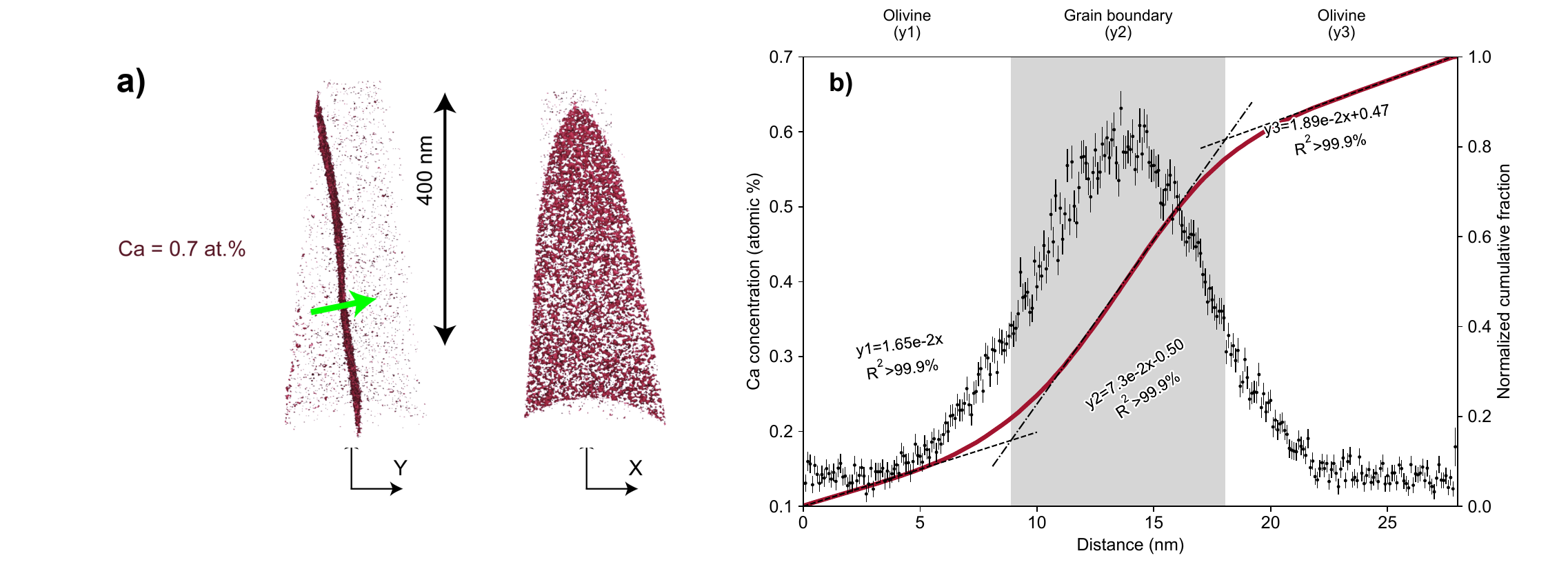}
\caption*{\textbf{Figure 4.} Calcium (Ca) interfacial enrichment at the R\'eunion olivine grain boundary and demonstration of the grain-boundary chemical-thickness measurement (APT). (a) Ca iso-concentration surface at 0.7 at.\%, shown as two orthogonal projections of the reconstruction; the Ca enrichment delineates a well-defined, planar grain boundary, the green arrow indicates the direction along which the 1-D compositional profile (Figure 5) is extracted, and the axis triads give the spatial orientation of each projection. (b) Demonstration of the GB chemical-thickness measurement using the Ca distribution: the measured Ca concentration (black) is plotted against distance across the boundary with the GB region shaded in gray, and the normalized cumulative-fraction curve (red) is fitted with three linear segments corresponding to the two olivine interiors (y1, y3) and the GB region (y2); the GB chemical thickness is taken as the distance between the intersections of the fitted lines. The corresponding Al iso-surface for this boundary, together with the Ca and Al iso-surfaces for the Hawaiian sample, are shown in Figure S6.}
\end{figure}

Figure 5 and Figure S7 present the one-dimensional compositional profiles across GBs in the Réunion and Hawaiian olivine samples, respectively. In Réunion, total elemental Fe is 6.15 at.\% in the matrix and 6.51 at.\% as the $\pm2$ nm boundary mean (ratio 1.06), consistent with Table 1 and showing only a modest increase; Mg decreases while Si remains nearly constant. Over the same boundary window, Ca averages 0.54 at.\% against 0.146 at.\% in the matrix (ratio 3.71), and Al averages 0.45 against 0.158 at.\% (ratio 2.83). P, Ti and Na also increase at the interface.

Similarly, in Mauna Loa olivine, Fe increases from a 5.75 at.\% matrix mean to 6.14 at.\% in the $\pm2$ nm window (ratio 1.07). Ca averages 0.76 at.\% against 0.127 at.\% in the matrix (ratio 5.98), and Al averages 0.35 against 0.089 at.\% (ratio 3.98). Hydrogen shows an apparent peak in both samples; because APT hydrogen signals are commonly affected by residual-gas and field-evaporation artifacts, we do not interpret it as structural incorporation, and H is omitted from Figure 6. The Mauna Loa P, Ti and Na profiles show boundary enrichment, whereas the nominal Sm-ranged signal is only weakly elevated and is consistent with its comparatively homogeneous ion map (Figure S5).

\begin{figure}[htbp]
\centering
\includegraphics[width=0.95\textwidth]{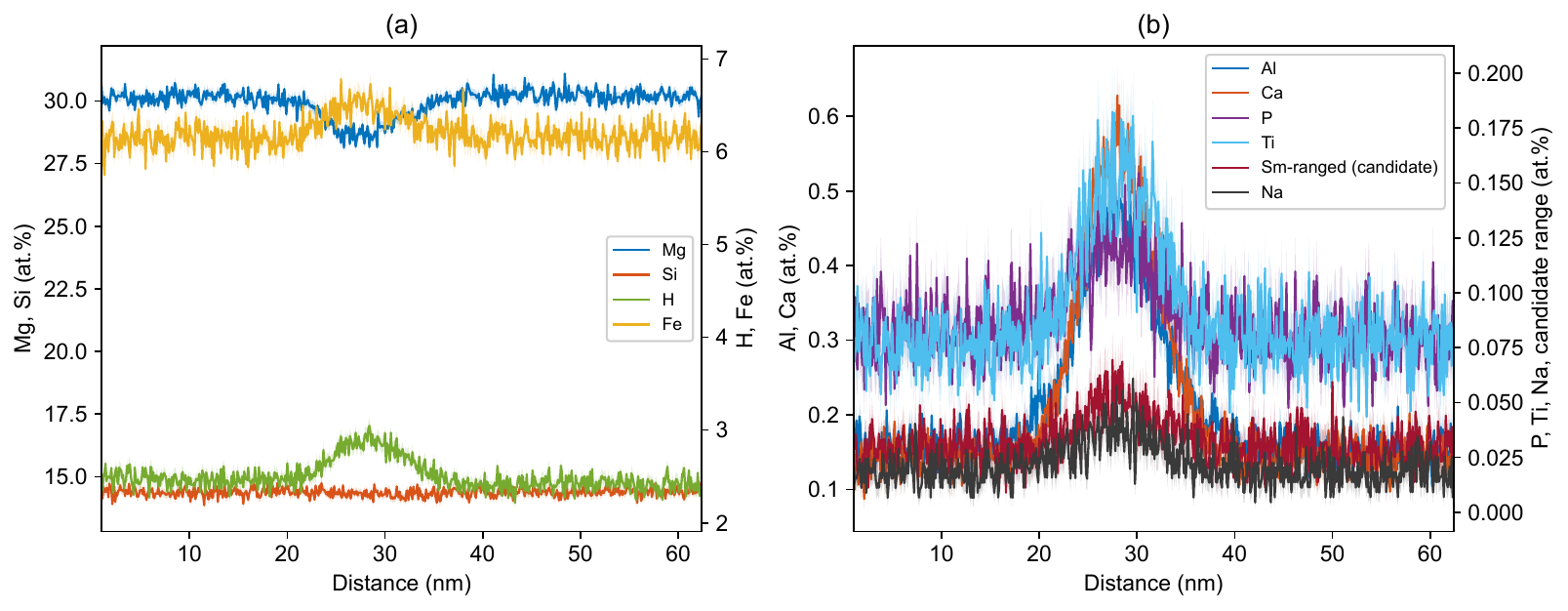}
\caption*{\textbf{Figure 5.} One-dimensional APT profile across the R\'eunion boundary, regenerated directly from reconstruction R13\_05409-v01, Cube ID 6, Y-axis (63.4 nm at 0.10 nm spacing), along the green arrow in Figure 4. The lateral Cube ID 6 dimensions are not retained in the AP Suite export; the ROI footprint is shown in Figure 4. Shading denotes reported $\pm1\sigma$ counting statistics. (a) Mg and Si (left axis) with H and total elemental Fe (right axis). Fe is the AP Suite elemental total, not a single Fe-ion channel; its matrix and $\pm2$ nm boundary means are 6.15 and 6.51 at.\%, respectively. H is artifact-prone and omitted from Figure 6. (b) Al and Ca (left axis) with P, Ti, Na and the nominal Sm-ranged candidate signal (right axis). Réunion point maxima are 0.63 at.\% Ca and 0.50 at.\% Al; the more robust $\pm2$ nm means are 0.54 and 0.45 at.\%. The Sm-ranged trace is not a confirmed elemental Sm measurement.}
\end{figure}

To assess the impact of deformation on GB chemical thickness, we compared our results from the undeformed Réunion and Hawaiian olivine samples with previously published data on deformed wehrlite (Cukjati et al., 2019) and deformed mylonite (Tacchetto et al., 2021) (Figure 6). The undeformed samples exhibit systematically larger GB thicknesses across most measured elements: for the most strongly enriched elements the GB width in the Réunion olivine reaches \textasciitilde{}10--15 nm. The broad Na value and the R\'eunion nominal SmO\textsuperscript{2+}-channel width are sensitive to the cumulative-fraction regression threshold and should be read as upper estimates, whereas Al, Ca, Ti and P give threshold-stable widths of \textasciitilde{}9--13 nm (Figure 6). No reliable Sm-channel width was determined for Hawaii, and Sm widths were not reported for the two comparator studies; these missing values are shown as n.d., not zero. In contrast, the deformed wehrlite and mylonite show significantly thinner GB enrichment zones, with most reported elements maintaining thicknesses below 5 nm. Taken at face value, this difference would suggest that deformation reduces the extent of element segregation at GBs (through dynamic recrystallization, dislocation-assisted transport, or removal of intergranular melt films). We caution, however, that the comparison is across different samples, laboratories, instruments, and reconstruction protocols, and that absolute chemical widths can be affected by APT local-magnification artifacts; combined with our single boundary per locality, the contrast should be regarded as suggestive rather than established (see Discussion).

\begin{figure}[htbp]
\centering
\includegraphics[width=0.7\textwidth]{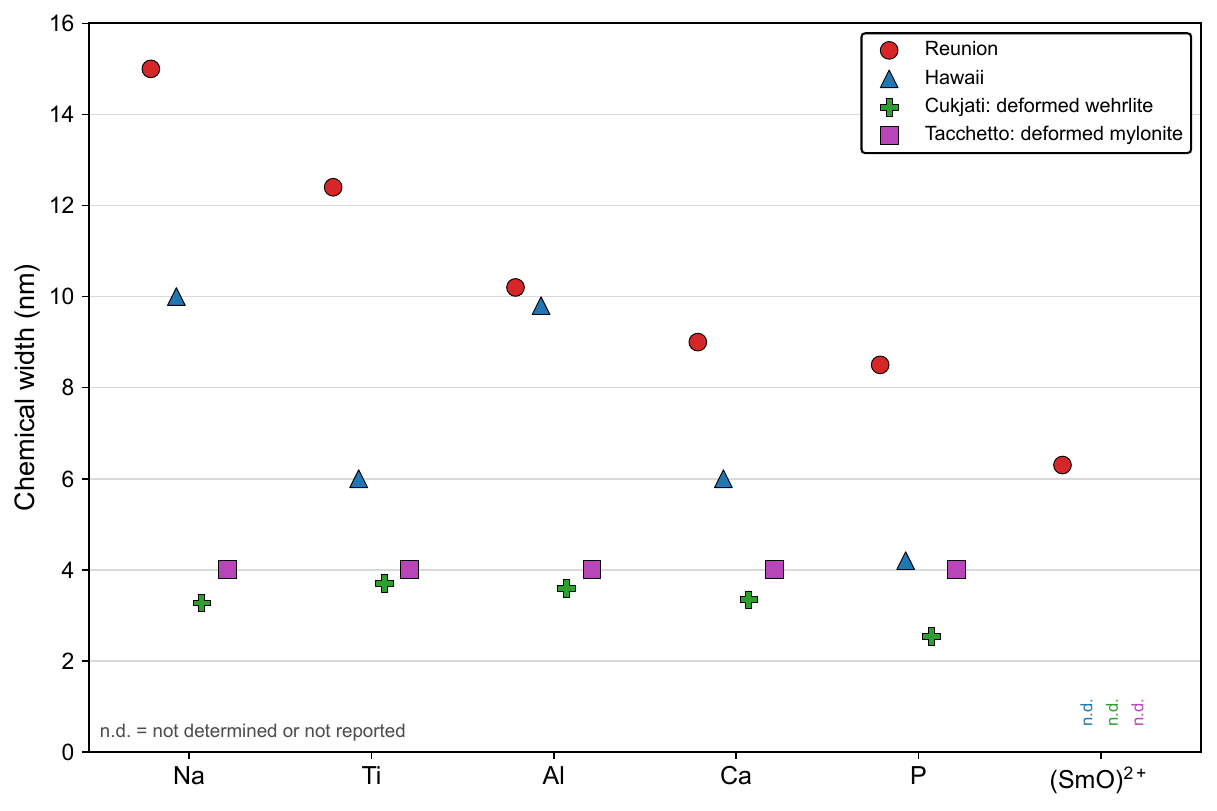}
\caption*{\textbf{Figure 6.} GB chemical widths for selected elements and the nominal (SmO)\textsuperscript{2+} mass channel in R\'eunion and Hawaiian olivine (this study), compared with deformed wehrlite (Cukjati et al., 2019) and deformed mylonite (Tacchetto et al., 2021). Points are not connected because the horizontal axis lists nominal chemical categories rather than a continuous variable. The broad Na value and the R\'eunion (SmO)\textsuperscript{2+}-channel width are sensitive to the cumulative-fraction regression threshold and should be read as upper estimates, whereas Al, Ca, Ti and P give threshold-stable widths. No reliable Sm-channel width was determined for Hawaii, and Sm widths were not reported by the comparator studies; these absent values are marked n.d. at the baseline and are not measured zeros. H is omitted as unreliable, reflecting APT hydrogen artifacts.}
\end{figure}

\subsubsection{Candidate rare-earth-related mass windows}
Rare-earth elements are difficult to identify in mafic APT spectra because nominal 2+/3+ atomic-ion, monoxide and dioxide positions in the \textasciitilde{}44--90 Da range overlap abundant Si\textsubscript{x}O\textsubscript{y} and Fe\textsubscript{x}O\textsubscript{y} matrix clusters, while the less-interfered singly charged REE region is flat background. We generated natural-abundance isotope templates and screened 112 nominal REE-related mass positions for spatial enrichment at the locally defined boundary (Supplementary Text S1; Table S2; Figures S8--S9). The spatial screen establishes whether a mass window is boundary associated; it does not identify the ion responsible.

The largest candidate-window response occurs at 83.96 Da, the nominal position of (\textsuperscript{152}Sm\textsuperscript{16}O)\textsuperscript{2+}, with a core/matrix density ratio of 1.35. However, the expected \textsuperscript{147}SmO\textsuperscript{2+}, \textsuperscript{149}SmO\textsuperscript{2+} and \textsuperscript{152}SmO\textsuperscript{2+} spacing and relative intensities are not resolved as a diagnostic envelope, and co-segregating molecular clusters involving independently enriched elements cannot be excluded. The nominal Sc-, Y- and Gd-related windows have the same identification limitation. We therefore report these only as candidate REE-related mass windows and make no elemental Sm, Sc, Y or Gd claim. The detected mass-to-charge species also cannot determine how any carrier was bonded or charge balanced before field evaporation.

\section{Discussion}

\subsection{Grain boundaries as nanoscale trace-element reservoirs}
A central result of this study is that olivine grain boundaries are not chemically equivalent to adjacent crystal interiors. In both the Réunion and Hawaiian samples, APT reveals planar enrichment of Ca, Al, Na, Ti and P, whereas major olivine-forming components such as Mg, Si, O, Ni, Cr and Co remain comparatively homogeneous away from the boundary. Several additional boundary-enriched mass windows coincide with nominal REE-ion positions, but their elemental carriers are unresolved and they are not used to support element-specific conclusions. This contrast indicates that the analyzed GBs represent localized trace-element reservoirs rather than simple extensions of the olivine lattice. We note at the outset that these conclusions rest on one fully analyzed grain boundary per locality; the systematics described below should therefore be read as well-characterized case studies rather than statistically representative population averages.

The selectivity of this enrichment points to grain-boundary segregation, rather than indiscriminate trapping, as the underlying control on which elements concentrate at the boundary. The major, well-fitted cations that occupy the olivine M-sites comfortably --- Mg, Fe, Ni, and Mn, all divalent and of similar size --- remain essentially homogeneous across the boundary, whereas the enriched species are precisely those that the olivine lattice accommodates poorly. In the framework established for mantle olivine by Hiraga et al. (2004), a grain boundary is a structurally open, distorted region that can host such mis-fitted ions at lower energy than the crystal interior. In this picture each element partitions between lattice and boundary according to an element-specific segregation free energy, $\Delta$G\textsubscript{seg}, with the most poorly fitted ions favored at the boundary. We stress that this framework predicts \emph{which} elements should preferentially occupy the boundary; \emph{whether} a given boundary attains its equilibrium occupancy is a separate question, addressed directly by the coverage measurements below.

For Ca, which is abundant enough to quantify reliably, the consistent enrichment statistic is the mean Ca concentration within $\pm2$ nm of the Ca-defined boundary centre divided by the $|d|\geq8$ nm matrix mean. This ratio is 3.71 for R\'eunion and 5.98 for Mauna Loa; the corresponding single-bin peak ratios are 4.36 and 7.85 and are reported separately because they are more noise sensitive. The Ca$^{+}$ proximity-density maximum of 1.84 in Figure S8 is not compared numerically because it uses a different ion channel, normalization and spatial statistic. Cumulative-fraction regression gives Ca chemical widths of 9 nm for R\'eunion and 6 nm for Mauna Loa; the earlier phrase ``6--8 nm halo'' was only a visual approximation and is replaced here by the measured 6--9 nm range.

Using the explicit M-site calculation defined in Methods, signed integration over the full profiles gives $\Gamma_{\mathrm{Ca}}=2.937\pm0.194$ atoms\,nm$^{-2}$ for R\'eunion and $2.966\pm0.055$ atoms\,nm$^{-2}$ for Mauna Loa. Division by 8.98 M sites\,nm$^{-2}$ gives equivalent coverages of $0.327\pm0.022$ and $0.330\pm0.006$ monolayers. Against $X_{GM}^{\mathrm{Ca}}=0.003978\pm0.000111$ and $0.003497\pm0.000043$, respectively, both points lie near the upper published 1523 K isotherm of Hiraga et al. (2004); the R\'eunion point lies slightly below it, corresponding qualitatively to a temperature at or above 1523 K within this model. We do not extrapolate a higher-temperature parameter curve or quote a precise temperature. The appropriate interpretation is therefore a model-equivalent segregation temperature close to basaltic olivine crystallization, not a measured closure thermometer. That agreement provides direct support for formation during crystal growth at near-magmatic temperature, although synneusis followed by sufficient high-temperature residence could produce a similar occupancy.

The equivalent coverage is well below a saturated monolayer and is distributed over the 6--9 nm chemical halo rather than confined to one atomic plane. This smooth profile may be intrinsically diffuse, but it could also record diffusional broadening of an initially sharper, approximately uniform monolayer-scale enrichment into the adjacent grains during cooling. APT local-magnification effects may add further apparent broadening (Vurpillot et al., 2000). Both processes can redistribute the profile while approximately conserving its integrated interfacial excess, so the coverage and the decoupling of the $\sim$1 nm structural core from the wider chemical halo are more robust than the exact profile shape.

\begin{figure}[htbp]
\centering
\includegraphics[width=0.72\textwidth]{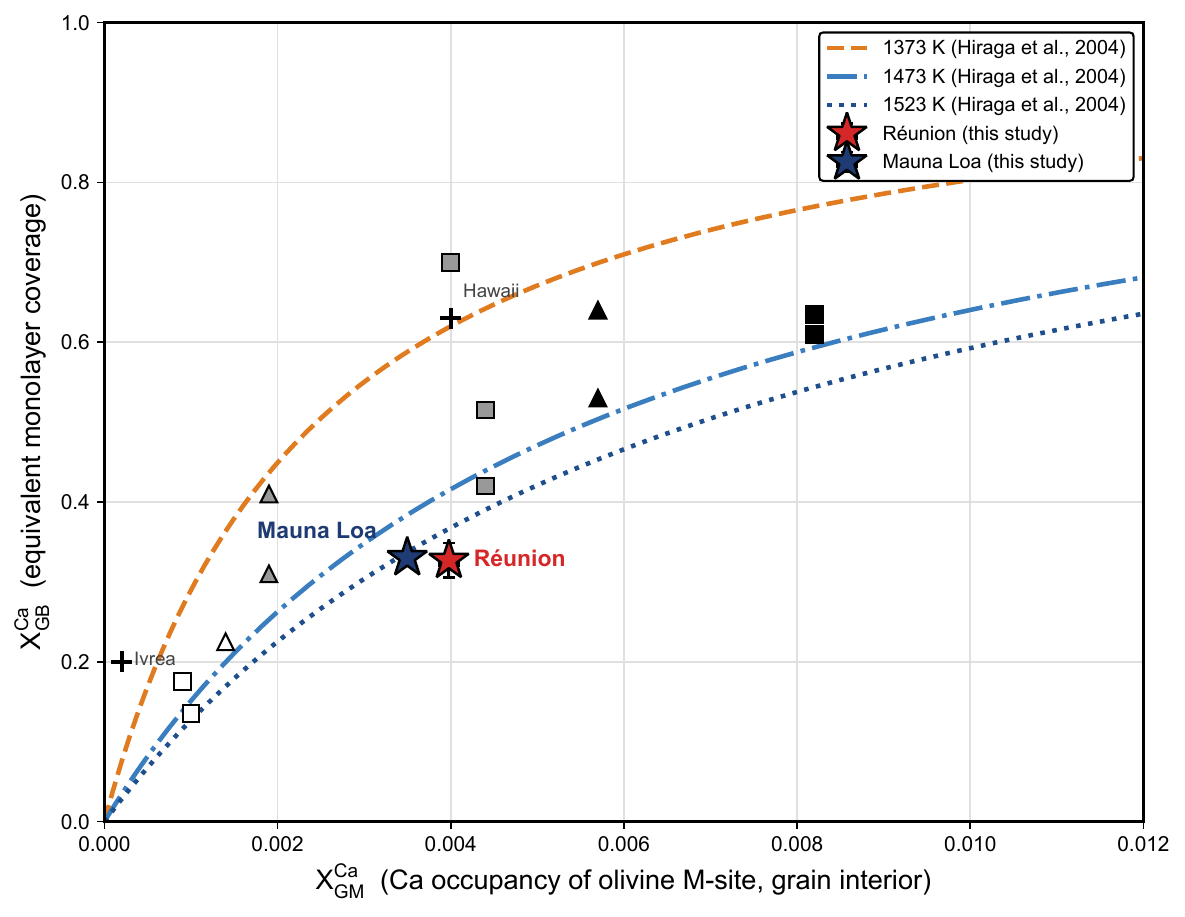}
\caption*{\textbf{Figure 7.} Grain-boundary Ca coverage versus grain-interior Ca M-site occupancy. Curves are calculated from equations 1--2 and the three published parameter sets (1373, 1473 and 1523 K) of Hiraga et al. (2004); the previously plotted 1300 and 1573 K extrapolations have been removed. Small symbols reproduce their digitized measurements. Stars are calculated directly from the deposited 1-D profiles using $X_{\mathrm{Ca}}=\mathrm{Ca}/(\mathrm{Mg}+\mathrm{Fe}+\mathrm{Ca}+\mathrm{Mn}+\mathrm{Ni}+\mathrm{Co})$, a matrix baseline at $|d|\geq8$ nm, signed integration over the full profile, $N_M=27.9$ nm$^{-3}$ and $N_M^{2/3}=8.98$ nm$^{-2}$. R\'eunion gives $\Gamma_{\mathrm{Ca}}=2.937\pm0.194$ atoms\,nm$^{-2}$ and $X_{GB}^{\mathrm{Ca}}=0.327\pm0.022$; Mauna Loa gives $2.966\pm0.055$ atoms\,nm$^{-2}$ and $0.330\pm0.006$. Error bars combine propagated counting statistics with sensitivity to baseline and profile-end choices; unquantified systematic APT bias is excluded. Both points lie near the upper published 1523 K curve, so the comparison supports near-crystallization-temperature segregation without requiring an extrapolated temperature curve.}
\end{figure}

Hiraga et al.'s synthetic aggregates were annealed at 1373--1523 K and quenched, whereas our boundaries occur in basaltic olivine that crystallized in a magma reservoir and was subsequently erupted. The two study points plot near the 1523 K end of Hiraga's published family, within the \textasciitilde{}1450--1570 K crystallization interval expected for Fo$_{83-84}$ olivine in these basalts (Villeneuve et al., 2008; Welsch et al., 2013; Putirka, 2008). This match provides support for co-growth of the grains and boundary at near-magmatic temperature. It is not unique proof: a synneusis boundary that remained hot long enough to approach the same occupancy could be indistinguishable. We therefore use ``model-equivalent segregation temperature,'' avoid an extrapolated numerical upper bound, and do not treat the nearly identical two study points as independent validation.

What sets $\Delta$G\textsubscript{seg}, and hence which elements enrich, is largely the misfit between each cation and the host M-site. To first order this misfit has an elastic component analogous to the lattice-strain model used for crystal--melt partitioning (Blundy \& Wood, 1994): a cation much larger or smaller than the optimal site radius strains the surrounding lattice and is preferentially expelled to the more compliant boundary. Plotting the segregation ratio against ionic radius makes this explicit (Figure 8). The isovalent divalent cations Ni, Mg, Fe, and Mn cluster near the site radius and show no enrichment, whereas the large isovalent Ca\textsuperscript{2+} is the most strongly enriched species and lies on the elastic strain curve; together the divalent cations trace the size-controlled trend (shaded band, Figure 8), the only valence our data constrain. Size misfit alone, however, cannot explain the small yet strongly enriched cations Al\textsuperscript{3+} and Ti\textsuperscript{4+}, which plot well above that curve.

The aliovalent cations require contributions to $\Delta$G\textsubscript{seg} that are not represented by the elastic size term alone. Al\textsuperscript{3+} and Ti\textsuperscript{4+} plot well above the Ca-anchored isovalent trend, but this departure cannot be assigned uniquely to charge: boundary coordination, bonding, chemical potentials, site availability, space charge, and coupled defects may all contribute. Local charge balance nonetheless requires aliovalent incorporation to be coupled to compensating defects or species. Na\textsuperscript{+} substituting on a divalent M-site could provide negative relative charge, whereas P\textsuperscript{5+} substituting for tetrahedral Si\textsuperscript{4+} adds positive relative charge and therefore cannot by itself compensate Al\textsuperscript{3+} or Ti\textsuperscript{4+} on an M\textsuperscript{2+} site. Their co-enrichment is consistent with coupled boundary defect chemistry, but the specific site occupancies and charge-balanced substitution reactions are not determined by the present APT data. Na also plots below the size-only curve, reinforcing that ionic radius alone is insufficient to describe aliovalent segregation; the sign and magnitude of the additional contribution, however, remain unresolved. As Hiraga et al. (2004) noted, the combination of size and charge resists a simple quantitative description for aliovalent ions, so we treat these departures qualitatively. Separate curves cannot be fitted for the aliovalent charge classes because the present dataset provides only one usable M-site cation for each class and does not constrain the space-charge and charge-compensation contributions to their segregation free energies. Figure 8 therefore supports a size-controlled description only for the isovalent divalent series and shows that additional, presently unconstrained terms are required for Al, Ti, and Na.

\begin{figure}[htbp]
\centering
\includegraphics[width=0.72\textwidth]{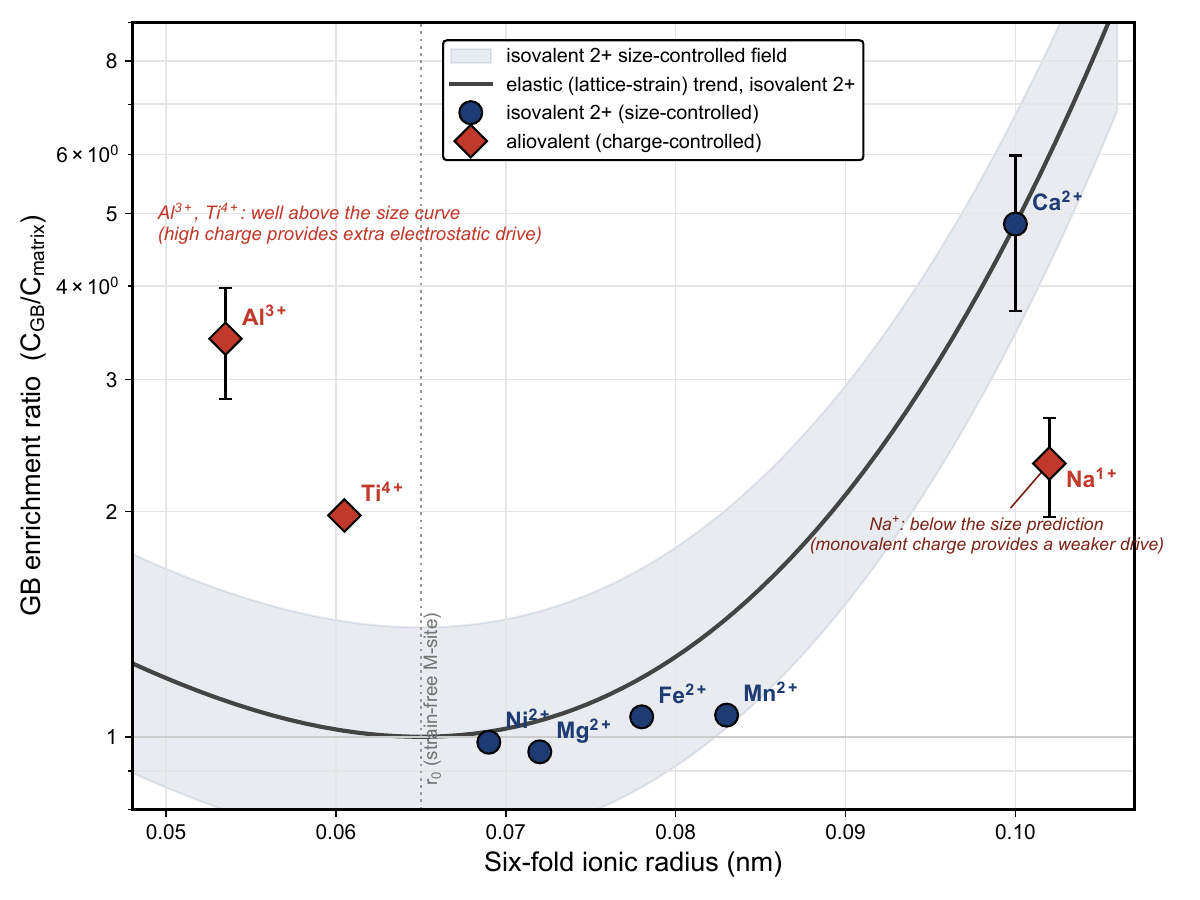}
\caption*{\textbf{Figure 8.} Grain-boundary enrichment ratio ($C_{\mathrm{GB}}/C_{\mathrm{matrix}}$, measured from the 1-D profiles as the mean of the two samples; error bars span R\'eunion and Mauna Loa) versus six-fold ionic radius. Ca has sample ratios of 3.71 and 5.98 and the plotted two-sample mean is 4.84. The solid curve is the elastic (lattice-strain) trend for isovalent M-site cations, from the misfit strain energy of Hiraga et al. (2004, their eq. 2; $r_0 = 0.065$ nm), anchored to Ca; the shaded band ($\pm$40\%) marks the corresponding isovalent-2+ size-controlled field, drawn as a guide to emphasize that only the 2+ cations are constrained by this trend. Isovalent divalent cations that fit the M-site (Ni, Mg, Fe, Mn) show essentially no enrichment (ratio $\approx$ 1) and sit at the base of the strain valley, whereas the large isovalent Ca\textsuperscript{2+} is the most strongly enriched and follows the curve. The aliovalent cations depart from this size-only trend: the small, highly charged Al\textsuperscript{3+} and Ti\textsuperscript{4+} plot well above it (an additional electrostatic, charge-compensation contribution), while the large but monovalent Na\textsuperscript{+} plots below it (a weaker drive than its size alone would predict). P\textsuperscript{5+} is also enriched (ratio $\approx$ 1.4) but substitutes for tetrahedral Si rather than the M-site, so it is not plotted on this M-site radius axis.}
\end{figure}

The combined TEM-APT approach further shows that the structural and chemical dimensions of these reservoirs are decoupled. TEM observations indicate that the crystallographic GB core is extremely narrow, with structural widths of approximately 1.15 nm in the Réunion sample and 1.05 nm in the Hawaiian sample. These boundaries are crystalline and show no evidence for continuous amorphous or glassy intergranular films. In contrast, APT concentration profiles show that chemical enrichment extends over wider regions. For several incompatible elements (Na, Ti, Al and P) the chemical width reaches several nanometers, locally \textasciitilde{}10--15 nm. Therefore, the GB should not be treated only as a zero-thickness crystallographic plane, but as a structurally narrow core surrounded by an element-specific chemical halo (Figure 9).

\begin{figure}[htbp]
\centering
\includegraphics[width=0.78\textwidth]{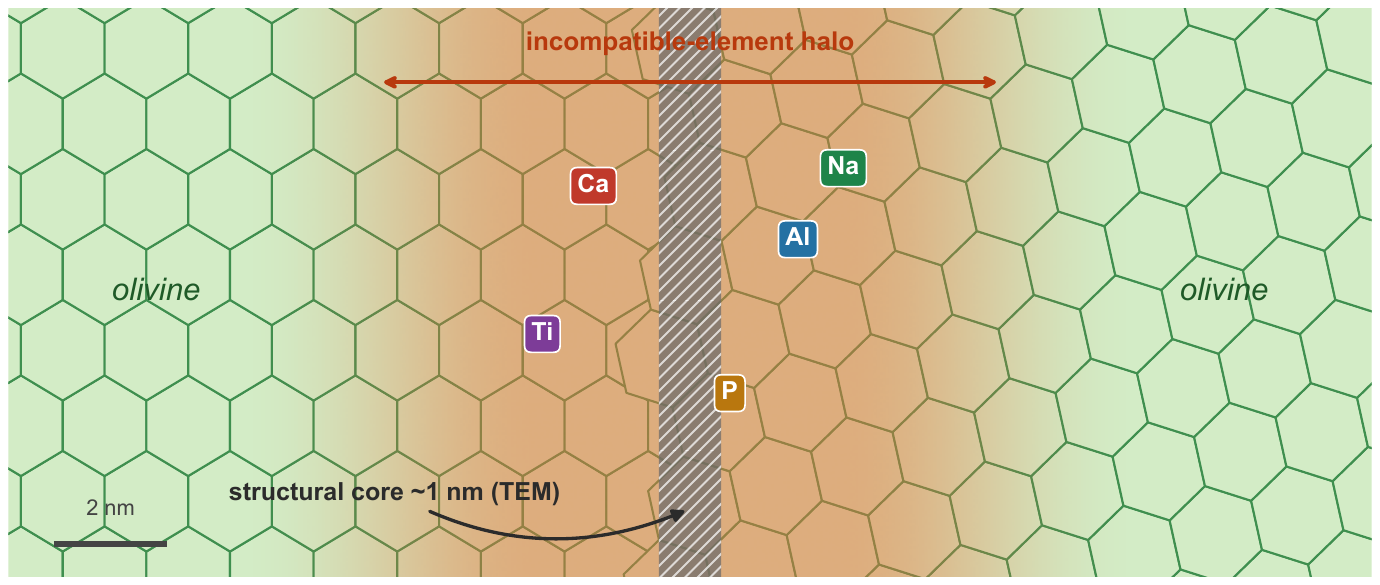}
\caption*{\textbf{Figure 9.} Schematic of an olivine--olivine grain boundary between two misoriented olivine grains (honeycomb lattices), illustrating the decoupling of its structural and chemical dimensions. The crystallographic core resolved by TEM is extremely narrow ($\sim$1 nm; central hatched band), whereas the incompatible-element enrichment forms a much broader, diffuse chemical halo (orange shading) that extends over several to $\sim$10--15 nm. Labeled chips mark the species this study finds enriched at the boundary --- Ca, Al, Ti, Na and P --- clustered near the core. The cartoon is qualitative; a single generic halo is drawn for clarity, but the chemical width is element-specific (widest for Na, Ti, Al and P; see Figure 6).}
\end{figure}

This distinction is important for separating true GB segregation from trapped melt. Residual melt may have supplied incompatible elements to the boundary during late-stage crystallization or cooling, but the final nanoscale distribution is not consistent with a simple continuous melt film. TEM confirms that the boundary is crystalline, and APT shows selective enrichment of specific incompatible elements rather than a wholesale melt-like composition. This is quantitatively borne out by the major-element framework, which a trapped basaltic film would strongly perturb but which our 1-D profiles leave essentially intact. The most obvious melt discriminator is Si: a basaltic liquid carries $\sim$30\% more Si than olivine on an atomic basis, yet across the boundary Si remains within $\sim$1\% of its matrix value (boundary/matrix $\approx$ 0.99 and 1.00 for R\'eunion and Mauna Loa, respectively) and O is likewise unchanged ($\approx$ 1.00). Fe, the element most plausibly mistaken for a melt signal, rises only marginally ($\approx$ 1.06--1.07) --- within counting scatter and far below the incompatible cations --- and the combined ferromagnesian framework Mg\,+\,Fe is preserved to within $\sim$2--3\% ($\approx$ 0.97--0.98); Fe behaves like the other isovalent M-site cations Mg, Mn and Ni, which show no enrichment (Figure 8). Because a basaltic melt has several-fold \emph{lower} Mg\,+\,Fe than olivine in atomic terms, a continuous melt film would raise Si and collapse the ferromagnesian content --- the opposite of what we observe. The enrichment is therefore restricted to the incompatible cations that misfit the olivine sites (Ca, Al, Ti, Na and P; boundary/matrix $\approx$ 1.4--6) superimposed on an otherwise unperturbed olivine framework, as expected for crystalline grain-boundary segregation rather than entrapped melt. We caution, however, that selectivity alone does not exclude a melt-derived contribution, because incompatible elements are also concentrated in residual melt; cleanly distinguishing in-boundary segregation from a former melt film or a melt-fed diffusion halo would require comparing the GB element ratios with those of the coexisting melt, which our single-boundary datasets do not allow. The diffuse, multi-nanometer halo --- especially for P --- is the component most likely to carry such a melt-related contribution. With that caveat, the boundaries are best interpreted as crystalline interfacial reservoirs that have interacted and partly equilibrated locally with the olivine lattice and surrounding melt, without preserving a continuous trapped glass.

A direct future test is to compare complete boundary-resolved trace-element patterns with those of the coexisting residual melt. Trapped melt should broadly retain the residual-liquid REE pattern and its associated major-element framework, whereas grain-boundary segregation should superimpose element-specific fractionation governed by ionic radius, charge, boundary structure, and finite site saturation. The trivalent REE series is particularly promising because variations across the lanthanides could reveal a radius-dependent segregation pattern that differs from the coexisting melt pattern. Paired APT measurements of grain boundaries and adjacent glass in experiments using terrestrial and lunar analog compositions could be combined with charge-dependent lattice-strain models or molecular-dynamics calculations of segregation free energies. Ratios among elements with similar melt incompatibility but different predicted boundary affinities would then provide discriminants between trapped melt and interfacial storage. The present mass-spectrum overlaps preclude such a full REE-pattern test here, but this approach offers a quantitative way to separate the two reservoirs in future KREEP and lunar magma-ocean models.

In this framework, the trace-element budget of a polycrystalline olivine aggregate includes at least three reservoirs: olivine crystal interiors, GB chemical domains, and residual or interstitial melt. The volumetric contribution of GBs is small, but their enrichment factors can be large, especially for elements with low olivine lattice solubility. Consequently, GBs may contribute disproportionately to the apparent trace-element inventory of olivine aggregates. This effect becomes especially important for fine-grained or aggregate-rich samples, where the total GB area per unit volume is high.

\subsection{Effects of grain-boundary segregation: chemical and physical consequences}
The chemical effect of GB segregation is element-specific. Ca and Al show clear boundary-parallel enrichment in both the Réunion and Hawaiian samples. The raw 0.10-nm profiles give point maxima of 0.63 and 0.50 at.\% in Réunion and 1.00 and 0.45 at.\% in Mauna Loa for Ca and Al, respectively; because point maxima are noise-sensitive, the quantitative comparison uses the $\pm 2$~nm boundary means reported above. Phosphorus also shows strong boundary-related enrichment, consistent with its limited compatibility in the olivine lattice and its sensitivity to rapid growth, late-stage melt evolution, and interface-controlled redistribution.

These chemical effects matter because they change how the trace-element inventory of an olivine aggregate should be interpreted. Elements with broad chemical halos (such as Na, Ti, Al and P) may contribute disproportionately to the apparent trace-element budget even though the crystallographic GB core is only about 1 nm wide. Hydrogen-bearing species also peak at the GB in the APT profiles, but we do not interpret these as structural incorporation: APT hydrogen is strongly affected by analytical background, specimen preparation, and residual gas in the analysis chamber, and we therefore exclude H from our quantitative conclusions (Figure 6).

The physical effect of GB segregation is more indirect, but it is equally important. Segregation changes the local chemistry of an interface that already controls diffusion, grain-boundary mobility, grain-boundary sliding, melt connectivity, and aggregate cohesion in olivine-rich rocks (Hirth \& Kohlstedt, 2003; Dohmen \& Milke, 2010; von Bargen \& Waff, 1986). The enriched elements documented here may modify boundary energy, local bonding, defect density, and short-range diffusion pathways. We do not directly measure these properties in this study, so the mechanical consequences should be treated as implications rather than direct observations. Nevertheless, the data show that olivine GBs are not passive geometric surfaces. They are chemically modified interfaces whose composition may influence how olivine aggregates respond to melt interaction, cooling, and deformation.

One possible consequence is solute-assisted grain-boundary premelting. A grain boundary can become structurally disordered or develop a thin liquid-like film below the bulk melting temperature when that state lowers the total interfacial free energy (cf.\ Cantwell et al., 2014). Segregation of Ca, Al, P, Na, Ti, and REE to an olivine boundary changes its local composition, may lower its effective solidus, and may stabilize a disordered interfacial state. The chemically enriched crystalline boundaries observed here could therefore represent low-temperature precursors to premelting: with increasing temperature or incompatible-element activity, the same interface might undergo a transition from a crystalline boundary with a diffuse chemical halo to a disordered or liquid-like film. Such a mechanism would not be the sole cause of premelting, which can also arise from interfacial-energy effects in chemically simpler boundaries, but segregation could shift the temperature and thickness of the premelted state.

The present samples do not demonstrate premelting. TEM resolves crystalline boundaries without a continuous amorphous or glassy film, and all structural observations were made after cooling. Conversely, the absence of a quenched film does not exclude a liquid-like boundary at magmatic temperature, because a high-temperature interfacial film could crystallize during cooling while leaving part of its chemical enrichment as a diffuse halo. A direct test would combine controlled-temperature annealing or heating experiments with rapid quenching and correlated TEM--APT measurements to track boundary structure and composition as a function of temperature. Molecular-dynamics calculations using the measured enrichment could independently test whether the segregated composition lowers the temperature at which the olivine boundary disorders.

By analogy, recent three-dimensional studies of nanolites in basaltic scoria show that nanoscale crystallization can modify magma rheology not only by increasing crystal volume fraction but also by chemically differentiating the adjacent melt into higher-viscosity boundary layers (Bamber et al., 2025). If GB segregation similarly concentrates or removes components such as Ca, Al, P, alkalis, or hydrogen-bearing species at interfaces, it could generate local composition gradients --- between crystal surface, GB chemical halo, and residual melt --- that alter melt polymerization, charge balance, volatile speciation, and interfacial mobility. Whether such gradients are large enough to influence aggregate-scale rheology cannot be assessed from our data, since the effect would scale with the interfacial area per unit volume, the composition--viscosity sensitivity of the components, and how far the boundary layer is immobilized relative to the melt --- none of which we constrain. We therefore raise it only as a qualitative possibility, whose quantification will require boundary-layer viscosities and interfacial areas beyond the present scope.

The comparison with deformed olivine (Results; Figure 6) bears on this physical dimension. The broader chemical widths of our undeformed boundaries, relative to published APT studies of deformed wehrlite and mylonitic olivine, suggest that deformation can modify both the chemical and physical state of olivine GBs: broad halos preserved from crystallization and late-stage melt interaction may be narrowed, redistributed, or partially erased during deformation by dynamic recrystallization, boundary migration, enhanced diffusion, and dislocation-assisted transport. As cautioned in the Results, this contrast --- across different samples, laboratories, and reconstruction protocols, and with a single boundary per locality --- is suggestive rather than statistically established. Even so, it indicates that GB segregation is coupled to the physical evolution of the boundary, not only a chemical phenomenon.

\subsection{Implications for more evolved systems: pressure and bulk-composition controls}
The significance of GB trace-element storage may increase as magmatic systems become more evolved. During fractional crystallization, residual melts become progressively enriched in incompatible elements (McIntire, 1963). Olivine crystallizing or aggregating in contact with such melts may therefore develop stronger GB enrichments than olivine in more primitive melts, even if the olivine lattice remains a poor host for these elements. In this sense, GBs provide a pathway by which evolved melt signatures can be retained within nominally primitive olivine-rich aggregates.

A simple predictive framework can be written in terms of three factors: the concentration of element i in the residual melt, the tendency of that element to segregate to the GB, and the amount of GB material sampled by the analysis. In first-order form,
\[ C_i^{\mathrm{GB}} = K_i^{\mathrm{GB/melt}}(P,T,X_{\mathrm{melt}})\, C_i^{\mathrm{melt}}, \]
where $K_i^{\mathrm{GB/melt}}$ is an effective GB--melt segregation coefficient that depends on pressure, temperature, melt composition, and boundary character. The apparent crystal-side concentration measured in a polycrystalline aggregate can then be approximated as
\[ C_i^{\mathrm{app}} = C_i^{\mathrm{ol}} + \frac{A_{\mathrm{GB}}}{V}\, w_{c,i}\,\bigl(C_i^{\mathrm{GB}} - C_i^{\mathrm{ol}}\bigr), \]
where $A_{\mathrm{GB}}/V$ is the grain-boundary area per unit volume and $w_{c,i}$ is the element-specific chemical width. This expression emphasizes that even a narrow structural boundary can have a measurable effect when the chemical halo is broad, the boundary area is large, or the element is strongly enriched relative to the olivine lattice. We emphasize, however, that $K_i^{\mathrm{GB/melt}}$ is not constrained by our data --- although the Ca coverages discussed above are consistent with near-equilibrium segregation at the crystallization temperature --- so this framework is meant to organize the controlling variables rather than to predict absolute concentrations.

Pressure may also modulate GB enrichment. If the segregation free energy carries a volume term ($K_i \propto \exp(-\Delta G_i^{\mathrm{seg}}/RT)$ with $\Delta G_i^{\mathrm{seg}}$ including $P\,\Delta V_i^{\mathrm{seg}}$; cf.\ Lej\v{c}ek, 2010), compression could either suppress segregation (if it requires excess interfacial free volume) or enhance it (if it stabilizes denser, charge-balanced boundary configurations), so the effect need not be universal and should depend on each element's partial molar volume, charge, and bonding. Absent pressure-resolved data, its sign and magnitude remain an open hypothesis.

Bulk composition exerts a more direct control through the composition of the residual melt. More evolved or enriched melts, especially those with higher P, Ti, Al, Ca, Na, alkalis, volatiles, and REE contents, should increase the supply of elements available for GB segregation. Composition also affects melt structure and charge balance (Mysen \& Richet, 2019). For example, enrichment of high-field-strength or highly charged species may require coupled association with charge-balancing cations such as Ca, Na, or H-bearing species. Similarly, changes in melt polymerization, water content, oxygen fugacity, and Fe-Mg-Ti chemistry may modify both the activity of trace elements in the melt and the stability of boundary-localized complexes.

This framework leads to testable predictions. First, olivine aggregates crystallized from more evolved or incompatible-element-rich melts should show stronger GB enrichment and, for some elements, broader chemical widths. Second, samples with smaller grain size or higher aggregate density should have a larger GB contribution to apparent trace-element budgets because $A_{\mathrm{GB}}/V$ is larger. Third, pressure series experiments at fixed melt composition could determine whether GB enrichment is suppressed or enhanced by compression, thereby constraining the sign and magnitude of $\Delta V_i^{\mathrm{seg}}$. Finally, composition series experiments at fixed pressure and temperature could separate the effects of residual melt supply from the intrinsic tendency of each element to partition into the GB region.

Thus, the reservoir effect documented here in basaltic olivine aggregates may be even more important in evolved, crystal-rich, or rapidly crystallizing magmas where trace-element budgets are controlled by a combination of lattice partitioning, residual melt chemistry, nanoscale crystallization, and interfacial storage. In high-silica evolved systems, APT has likewise resolved incompatible-element redistribution at the nanometer scale --- for example, zircon--coffinite--xenotime solid solution concentrating REE, P, Zr, U and Th among accessory phases (Zhao et al., 2023) --- consistent with nanoscale partitioning of incompatible elements being a general feature of evolved magmatic systems. The present data do not yet define a quantitative pressure- or composition-dependent GB partitioning law, but they provide the observational basis for developing one.

An important end-member application of this framework is the lunar magma-ocean (LMO) cumulate pile. We single out the LMO not because our terrestrial olivine bears directly on the Moon, but because magma-ocean differentiation is thought to be a near-universal early stage of rocky-planet evolution (Elkins-Tanton, 2012), so an interfacial reservoir documented in terrestrial olivine carries broad relevance to rocky bodies. The lunar case is simply its best-characterized example: its cumulate stratigraphy and urKREEP framework are comparatively well established and stable, providing a clean, well-posed setting against which an additional nanoscale reservoir can be assessed. We use it here only to illustrate a possible implication; the actual interfacial segregation under lunar-specific pressure, temperature, and bulk-composition conditions is not constrained by our two terrestrial boundaries and would require dedicated experiments and analyses. In classical LMO models (Snyder et al., 1992; Hess \& Parmentier, 1995; Charlier et al., 2018; Rapp \& Draper, 2018), the last residual liquid is represented by urKREEP (Warren \& Wasson, 1979; Warren, 1985), enriched in K, REE, P, Th, U, and other incompatible components, and its distribution is commonly evaluated in terms of trapped intercumulus liquid, late-stage KREEP-rich layers, or accessory phases. The present results suggest an additional reservoir that should be considered: a fraction of the urKREEP-like inventory may be retained as nanoscale interfacial enrichment along grain and phase boundaries within cumulates. This does not require GBs to be the dominant urKREEP host. Rather, it implies that cumulate layers with nominally low lattice partition coefficients can still acquire a measurable KREEP component if they crystallized, aggregated, or annealed in contact with evolved residual liquid. Because the residual liquid becomes progressively more enriched in incompatible elements as crystallization proceeds, the absolute interfacial enrichment is expected to intensify upward through the cumulate pile, potentially approaching saturation in the most evolved (ilmenite-bearing cumulate, IBC, and urKREEP) layers --- a testable prediction summarized schematically in Figure 10.

\begin{figure}[htbp]
\centering
\includegraphics[width=\textwidth]{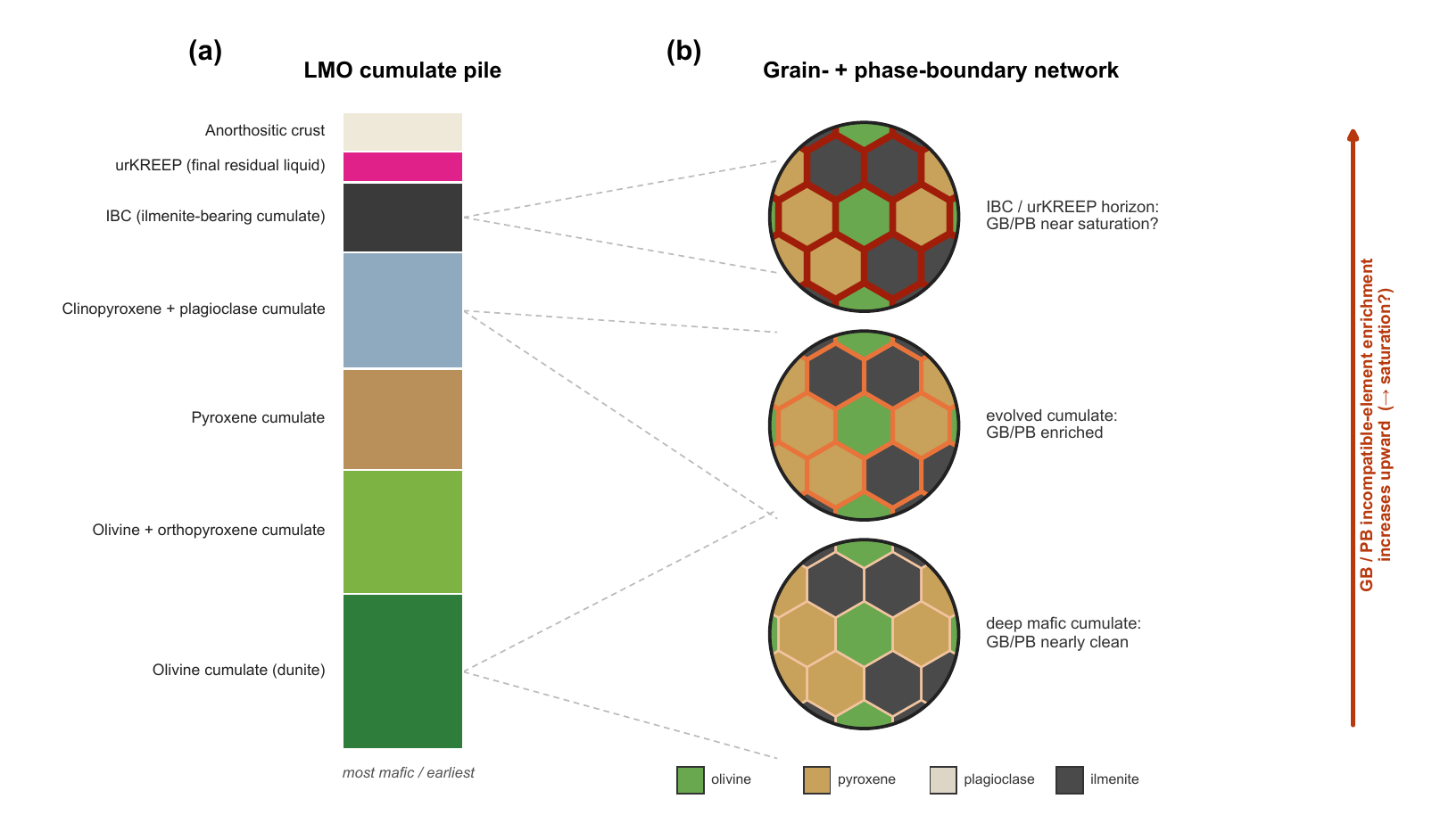}
\caption*{\textbf{Figure 10.} Highly schematic summary of the hypothesis (to be tested) that the grain- plus phase-boundary (GB\,+\,PB) network is a distributed incompatible-element reservoir in the lunar magma-ocean (LMO) cumulate pile. \textbf{(a)} Stratigraphic column of LMO cumulates, from the most mafic, earliest cumulates (bottom) through pyroxene and clinopyroxene\,+\,plagioclase cumulates to the ilmenite-bearing cumulate (IBC), urKREEP and anorthositic crust (top); the arrow marks increasing crystallization, melt evolution, and incompatible-element enrichment. \textbf{(b)} The GB\,+\,PB network sampled at three stratigraphic levels (different colors denote different cumulus minerals): as the residual liquid evolves, the \emph{absolute} incompatible-element enrichment of the boundary network is hypothesized to intensify upward (boundaries shading from pale orange to deep red and thickening), possibly approaching saturation in the most evolved IBC/urKREEP layers. The relative shares carried by interfaces, trapped melt, and accessory phases remain unconstrained and are not predicted by this schematic.}
\end{figure}

This application is also motivated by recent petrogenetic studies of Chang'e-5 and Chang'e-6 lunar basalts. In the Chang'e-5 Nature study, the modeled mantle source was described as an 86\% crystallized LMO cumulate with 2\% trapped instantaneous residual liquid (TIRL), which was needed to reproduce the source Sr-Nd isotope systematics and incompatible-element characteristics (Tian et al., 2021). In the Chang'e-6 Nature study of South Pole-Aitken basin basalts, several LMO cumulate models likewise required small amounts of TIRL, about 0.3-1.0\%, to reproduce the measured source Sm/Nd ratios before evaluating partial melting and fractional crystallization (Zhou et al., 2025). These studies do not address grain boundaries directly, but they show that the geochemical interpretation of lunar cumulates already depends on minor residual-liquid reservoirs. The GB reservoir considered here provides a nanoscale form of the same mass-balance problem: some incompatible-element inventory attributed to trapped melt may be retained instead, or additionally, along grain and phase boundaries within the cumulate pile.

The abundance of urKREEP components in individual LMO cumulate layers can therefore be evaluated using a reservoir-resolved mass balance rather than mineral-melt partition coefficients alone. For a given layer, the total concentration of an incompatible element can be written conceptually as
\[ C_i^{\mathrm{layer}} = \sum_m \phi_m C_i^{m} + \phi_{\mathrm{tr}} C_i^{\mathrm{tr}} + f_{\mathrm{GB},i}\, C_i^{\mathrm{GB}}, \]
where $\phi_m$ and $C_i^{m}$ are the modal abundance and lattice concentration of each cumulus mineral, $\phi_{\mathrm{tr}} C_i^{\mathrm{tr}}$ represents trapped or intercumulus melt, and $f_{\mathrm{GB},i} \approx (A_{\mathrm{GB}}/V)\, w_{c,i}$ for the element-specific grain-boundary reservoir. In practice, $A_{\mathrm{GB}}/V$ can be estimated from grain size and texture, $w_{c,i}$ and $C_i^{\mathrm{GB}}$ from APT-TEM measurements or analog experiments, and $C_i^{\mathrm{tr}}$ from LMO crystallization models. This framework predicts that late, fine-grained, evolved, or accessory-mineral-rich cumulates should record the strongest urKREEP-like signatures, whereas deep early cumulates should show smaller interfacial contributions unless high pressure stabilizes boundary segregation. Comparing the calculated GB, trapped-melt, and accessory-mineral terms across stratigraphic layers would provide a way to test whether urKREEP is concentrated only in a late residual horizon or partly distributed through the cumulate pile as nanoscale interfacial storage.

We deliberately stop short of a numerical lunar mass balance: with only two terrestrial boundaries, propagating our measurements into specific urKREEP percentages would over-extrapolate beyond what the data support. Figure 10 therefore illustrates only the hypothesis that absolute interfacial enrichment may rise as residual liquid evolves. The fractional shares carried by grain and phase boundaries, trapped melt, and accessory phases cannot be ranked from the present data because each reservoir is controlled by different and presently unconstrained capacities, modal abundances, and partitioning relations. Whether nanoscale interfacial storage is quantitatively comparable to the small trapped-melt fractions (sub-per-cent to per-cent) invoked in source models of Chang'e-5 and Chang'e-6 basalts remains an open and testable question that requires lunar-specific measurements rather than extrapolation from terrestrial olivine.

\section{Conclusions}
This study provides a nanoscale structural and chemical characterization of grain boundaries (GBs) in undeformed basaltic olivine from Piton de la Fournaise (La Réunion) and Mauna Loa (Hawaii), integrating EBSD, TEM, and atom probe tomography (APT). Our principal findings are as follows:

(1) Crystalline boundaries enriched in incompatible elements. TEM shows that the olivine GBs are crystalline and only \textasciitilde{}1.05--1.15 nm wide, with no continuous glassy or amorphous intergranular film resolved, while APT reveals systematic enrichment of Ca, Al and P together with Na and Ti. Several other enriched mass windows coincide with nominal REE-ion positions, but their elemental carriers are unresolved and no Sm, Sc, Y or Gd detection is claimed. No continuous trapped glass film is preserved, but the data do not exclude a former melt film, a crystallized intergranular liquid, or a melt-fed diffusion halo.

(2) Size-dependent segregation with additional aliovalent contributions, consistent with equilibrium two-dimensional Ca partitioning. Cations that fit the olivine M-site (Mg, Fe, Ni, Mn) remain homogeneous across the boundary, whereas large Ca follows the isovalent size-misfit trend. Al, Ti, and Na depart from that size-only relation, showing that additional contributions from boundary coordination, bonding, site availability, space charge, or coupled defects are required; the specific site occupancies and charge-balanced substitutions, including the origin of P co-enrichment, are not resolved. The measured Ca coverages (0.327--0.330 equivalent monolayers) lie near the upper published 1523 K isotherm of Hiraga et al. (2004), close to the basaltic olivine crystallization temperature. This agreement supports boundary formation during crystal growth, although synneusis followed by sufficient high-temperature residence could produce a similar signature; the Ca is distributed over cumulative-fraction widths of \textasciitilde{}6--9 nm rather than a saturated single-plane monolayer.

(3) A narrow structural core with a wider chemical halo, and a three-reservoir budget. The crystallographic core is only \textasciitilde{}1 nm wide, but the chemical enrichment extends over several nanometers and locally reaches \textasciitilde{}10--15 nm for the most strongly enriched elements. The trace-element budget of a polycrystalline olivine aggregate therefore comprises three reservoirs --- crystal interiors, grain boundaries, and residual melt; although GBs occupy a tiny volume, their large enrichment factors can contribute non-negligibly to the apparent crystal-side inventory and should be accounted for when interpreting apparent olivine--melt partitioning.

(4) Sensitivity to deformation and significance for evolved and planetary systems. The undeformed boundaries studied here preserve broader chemical widths than published deformed olivine, tentatively suggesting --- across differing samples and protocols --- that deformation narrows or redistributes GB segregation. Because the boundary's share of the trace-element inventory grows with melt evolution and with decreasing grain size, this interfacial reservoir should be most important in evolved, crystal-rich, or fine-grained cumulates, and may contribute qualitatively to the incompatible-element budget of lunar magma-ocean cumulates --- part of an inventory conventionally attributed to trapped melt --- though we leave its lunar magnitude as an open, testable question rather than a numerical estimate.

Together, these results establish olivine grain boundaries as minor but stable nanoscale reservoirs of confirmed incompatible elements, and show that resolving their chemistry --- now achievable with APT when mass assignments are interference constrained --- is necessary for accurate trace-element mass balance in olivine-rich terrestrial and planetary systems. We emphasize, however, that these conclusions rest on one fully analyzed boundary per locality; the systematics proposed here --- the size--charge selectivity, the near-crystallization model-equivalent Ca segregation temperature, and the deformation contrast --- should be regarded as well-characterized case studies and tested against a larger population of boundaries spanning grain-boundary character, composition, and cooling history.

\section*{Acknowledgments}
This work was supported by the U.S. National Science Foundation, Division of Earth Sciences, under award EAR-1947439 (Petrology and Geochemistry Program) to R.C. and S.P., and through grants EAR-1144668 (Petrology and Geochemistry Program) and EAR-1620474 (Geophysics Program). We thank Beno\^it Welsch for providing the Piton de la Fournaise (La R\'eunion) olivine clots and Fred Anderson for the Hawaiian olivine clots. EBSD, focused-ion-beam, TEM and atom-probe analyses were carried out at the Center for Nanoscale Systems (CNS), Harvard University, a member of the National Nanotechnology Coordinated Infrastructure Network (NNCI), supported under NSF award no.\ ECCS-1541959. We also thank Dr.\ Jules Gardener for the TEM analyses and Timothy Cavanaugh for the EBSD analyses.

\section*{Author contributions}
R.C. and S.P. conceived and designed the project and acquired funding. W.Z. and A.A. performed the experimental analyses (focused-ion-beam specimen preparation, TEM, and atom probe tomography). W.Z. conceived the framing of the manuscript, analyzed and interpreted the data, prepared the figures, and wrote the original draft. R.C., S.P., and G.H. contributed to the interpretation and discussion. All authors reviewed and edited the manuscript.

\section*{Data availability}
The atom probe tomography, TEM, and EBSD datasets supporting the findings of this study are available from the corresponding author on reasonable request. Upon acceptance, or earlier if requested during review, the datasets will be deposited in a publicly accessible repository (the Brown University Digital Repository, https://repository.library.brown.edu).

\section*{Declaration of competing interest}
The authors declare that they have no known competing financial interests or personal relationships that could have appeared to influence the work reported in this paper.

\section*{References}
Azevedo, S. \& Nespolo, M. (2017) Twinning in olivine group revisited. European Journal of Mineralogy 29, 213--226. \url{https://doi.org/10.1127/ejm/2017/0029-2598}

Bamber, E. C. et al. (2025) 3D quantification of nanolites using X-ray ptychography reveals syn-eruptive nanocrystallisation impacts magma rheology. Nature Communications 16, 7083. \url{https://doi.org/10.1038/s41467-025-62444-z}

Blundy, J. \& Wood, B. (1994) Prediction of crystal--melt partition coefficients from elastic moduli. Nature 372, 452--454. \url{https://doi.org/10.1038/372452a0}

Blundy, J. \& Wood, B. (2003) Partitioning of trace elements between crystals and melts. Earth and Planetary Science Letters 210, 383--397. \url{https://doi.org/10.1016/S0012-821X(03)00129-8}

Cantwell, P. R., Tang, M., Dillon, S. J., Luo, J., Rohrer, G. S. \& Harmer, M. P. (2014) Grain boundary complexions. Acta Materialia 62, 1--48. \url{https://doi.org/10.1016/j.actamat.2013.07.037}

Charlier, B., Grove, T. L., Namur, O. \& Holtz, F. (2018) Crystallization of the lunar magma ocean and the primordial mantle--crust differentiation of the Moon. Geochimica et Cosmochimica Acta 234, 50--69. \url{https://doi.org/10.1016/j.gca.2018.05.006}

Cukjati, J. T., Cooper, R. F., Parman, S. W., Zhao, N., Akey, A. J. \& Laiginhas, F. A. T. P. (2019) Differences in chemical thickness of grain and phase boundaries: an atom probe tomography study of experimentally deformed wehrlite. Physics and Chemistry of Minerals 46, 845--859. \url{https://doi.org/10.1007/s00269-019-01045-x}

Dohmen, R. \& Milke, R. (2010) Diffusion in polycrystalline materials: grain boundaries, mathematical models, and experimental data. Reviews in Mineralogy and Geochemistry 72, 921--970. \url{https://doi.org/10.2138/rmg.2010.72.21}

Elkins-Tanton, L. T. (2012) Magma oceans in the inner solar system. Annual Review of Earth and Planetary Sciences 40, 113--139. \url{https://doi.org/10.1146/annurev-earth-042711-105503}

Famin, V., Welsch, B., Okumura, S., Bachèlery, P. \& Nakashima, S. (2009) Three differentiation stages of a single magma at Piton de la Fournaise volcano (Réunion hot spot). Geochemistry, Geophysics, Geosystems 10, Q01007. \url{https://doi.org/10.1029/2008GC002015}

Gault, B., Moody, M. P., Cairney, J. M. \& Ringer, S. P. (2012) Atom Probe Microscopy. Springer Series in Materials Science 160. Springer, New York. \url{https://doi.org/10.1007/978-1-4614-3436-8}

Gillot, P.-Y. \& Nativel, P. (1989) Eruptive history of the Piton de la Fournaise volcano, Réunion Island, Indian Ocean. Journal of Volcanology and Geothermal Research 36, 53--65. \url{https://doi.org/10.1016/0377-0273(89)90005-X}

Goldschmidt, V. M. (1937) The principles of distribution of chemical elements in minerals and rocks. Journal of the Chemical Society 1937, 655--673. \url{https://doi.org/10.1039/JR9370000655}

Hellman, O. C., Vandenbroucke, J. A., Rüsing, J., Isheim, D. \& Seidman, D. N. (2000) Analysis of three-dimensional atom-probe data by the proximity histogram. Microscopy and Microanalysis 6, 437--444. \url{https://doi.org/10.1007/S100050010051}

Hess, P. C. \& Parmentier, E. M. (1995) A model for the thermal and chemical evolution of the Moon's interior: implications for the onset of mare volcanism. Earth and Planetary Science Letters 134, 501--514. \url{https://doi.org/10.1016/0012-821X(95)00138-3}

Hiraga, T., Anderson, I. M. \& Kohlstedt, D. L. (2003) Chemistry of grain boundaries in mantle rocks. American Mineralogist 88, 1015--1019. \url{https://doi.org/10.2138/am-2003-0709}

Hiraga, T., Anderson, I. M. \& Kohlstedt, D. L. (2004) Grain boundaries as reservoirs of incompatible elements in the Earth's mantle. Nature 427, 699--703. \url{https://doi.org/10.1038/nature02259}

Hirth, G. \& Kohlstedt, D. (2003) Rheology of the upper mantle and the mantle wedge: a view from the experimentalists. In: Eiler, J. (ed) Inside the Subduction Factory. Geophysical Monograph 138, 83--105. American Geophysical Union, Washington, D.C. \url{https://doi.org/10.1029/138GM06}

Krakauer, B. W. \& Seidman, D. N. (1993) Absolute atomic-scale measurements of the Gibbsian interfacial excess of solute at internal interfaces. Physical Review B 48, 6724--6727. \url{https://doi.org/10.1103/PhysRevB.48.6724}

Lejček, P. (2010) Grain Boundary Segregation in Metals. Springer Series in Materials Science 136. Springer, Berlin Heidelberg. \url{https://doi.org/10.1007/978-3-642-12505-8}

Mancini, L. et al. (2014) Composition of wide bandgap semiconductor materials and nanostructures measured by atom probe tomography and its dependence on the surface electric field. The Journal of Physical Chemistry C 118, 24136--24151. \url{https://doi.org/10.1021/jp5071264}

Marquardt, K., Rohrer, G. S., Morales, L., Rybacki, E., Marquardt, H. \& Lin, B. (2015) The most frequent interfaces in olivine aggregates: the GBCD and its importance for grain-boundary related processes. Contributions to Mineralogy and Petrology 170, 40. \url{https://doi.org/10.1007/s00410-015-1193-9}

McIntire, W. L. (1963) Trace element partition coefficients---a review of theory and applications to geology. Geochimica et Cosmochimica Acta 27, 1209--1264. \url{https://doi.org/10.1016/0016-7037(63)90049-8}

McLean, D. (1957) Grain Boundaries in Metals. Clarendon Press, Oxford.

Meija, J., Coplen, T. B., Berglund, M., Brand, W. A., De Bièvre, P., Gröning, M., Holden, N. E., Irrgeher, J., Loss, R. D., Walczyk, T. \& Prohaska, T. (2016) Isotopic compositions of the elements 2013 (IUPAC Technical Report). Pure and Applied Chemistry 88, 293--306. \url{https://doi.org/10.1515/pac-2015-0503}

Michon, L., Lénat, J.-F., Bachèlery, P. \& Di Muro, A. (2015) Geology and morphostructural evolution of Piton de la Fournaise. In: Bachèlery, P., Lénat, J.-F., Di Muro, A. \& Michon, L. (eds) Active Volcanoes of the Southwest Indian Ocean: Piton de la Fournaise and Karthala. Active Volcanoes of the World. Springer, Berlin Heidelberg, pp. 45--59. \url{https://doi.org/10.1007/978-3-642-31395-0_4}

Mysen, B. O. \& Richet, P. (2019) Silicate Glasses and Melts, 2nd edn. Elsevier, Amsterdam.

Onuma, N., Higuchi, H., Wakita, H. \& Nagasawa, H. (1968) Trace element partition between two pyroxenes and the host lava. Earth and Planetary Science Letters 5, 47--51. \url{https://doi.org/10.1016/S0012-821X(68)80010-X}

Putirka, K. D. (2008) Thermometers and barometers for volcanic systems. Reviews in Mineralogy and Geochemistry 69, 61--120. \url{https://doi.org/10.2138/rmg.2008.69.3}

Rapp, J. F. \& Draper, D. S. (2018) Fractional crystallization of the lunar magma ocean: updating the dominant paradigm. Meteoritics \& Planetary Science 53, 1432--1455. \url{https://doi.org/10.1111/maps.13086}

Reddy, S. M., Saxey, D. W., Rickard, W. D. A., Fougerouse, D., Montalvo, S. D., Verberne, R. \& van Riessen, A. (2020) Atom Probe Tomography: Development and Application to the Geosciences. Geostandards and Geoanalytical Research 44, 5--50. \url{https://doi.org/10.1111/ggr.12313}

Rhodes, J. M. (1995) The 1852 and 1868 Mauna Loa picrite eruptions: clues to parental magma compositions and the magmatic plumbing system. In: Rhodes, J. M. \& Lockwood, J. P. (eds) Mauna Loa Revealed: Structure, Composition, History, and Hazards. Geophysical Monograph Series 92, 241--262. American Geophysical Union, Washington, D.C. \url{https://doi.org/10.1029/GM092p0241}

Ringwood, A. E. (1975) Composition and Petrology of the Earth's Mantle. McGraw-Hill, New York.

Salvany, T., Lahitte, P., Nativel, P. \& Gillot, P.-Y. (2012) Geomorphic evolution of the Piton des Neiges volcano (Réunion Island, Indian Ocean): competition between volcanic construction and erosion since 1.4 Ma. Geomorphology 136, 132--147. \url{https://doi.org/10.1016/j.geomorph.2011.06.009}

Saxey, D. W., Moser, D. E., Piazolo, S., Reddy, S. M. \& Valley, J. W. (2018) Atomic worlds: Current state and future of atom probe tomography in geoscience. Scripta Materialia 148, 115--121. \url{https://doi.org/10.1016/j.scriptamat.2017.11.014}

Schwindinger, K. R. \& Anderson, A. T. (1989) Synneusis of Kilauea Iki olivines. Contributions to Mineralogy and Petrology 103, 187--198. \url{https://doi.org/10.1007/BF00378504}

Snyder, G. A., Taylor, L. A. \& Neal, C. R. (1992) A chemical model for generating the sources of mare basalts: combined equilibrium and fractional crystallization of the lunar magmasphere. Geochimica et Cosmochimica Acta 56, 3809--3823. \url{https://doi.org/10.1016/0016-7037(92)90172-F}

Tacchetto, T., Reddy, S. M., Saxey, D. W., Fougerouse, D., Rickard, W. D. A. \& Clark, C. (2021) Disorientation control on trace element segregation in fluid-affected low-angle boundaries in olivine. Contributions to Mineralogy and Petrology 176, 59. \url{https://doi.org/10.1007/s00410-021-01815-3}

Thompson, K., Lawrence, D., Larson, D. J., Olson, J. D., Kelly, T. F. \& Gorman, B. (2007) In situ site-specific specimen preparation for atom probe tomography. Ultramicroscopy 107, 131--139. \url{https://doi.org/10.1016/j.ultramic.2006.06.008}

Tian, H.-C. et al. (2021) Non-KREEP origin for Chang'e-5 basalts in the Procellarum KREEP Terrane. Nature 600, 59--63. \url{https://doi.org/10.1038/s41586-021-04119-5}

Valley, J. W., Cavosie, A. J., Ushikubo, T., Reinhard, D. A., Lawrence, D. F., Larson, D. J., Clifton, P. H., Kelly, T. F., Wilde, S. A., Moser, D. E. \& Spicuzza, M. J. (2014) Hadean age for a post-magma-ocean zircon confirmed by atom probe tomography. Nature Geoscience 7, 219--223. \url{https://doi.org/10.1038/ngeo2075}

Villeneuve, N., Neuville, D. R., Boivin, P., Bachèlery, P. \& Richet, P. (2008) Magma crystallization and viscosity: a study of molten basalts from the Piton de la Fournaise volcano (La Réunion island). Chemical Geology 256, 242--251. \url{https://doi.org/10.1016/j.chemgeo.2008.06.039}

Vurpillot, F., Bostel, A. \& Blavette, D. (2000) Trajectory overlaps and local magnification in three-dimensional atom probe. Applied Physics Letters 76, 3127--3129. \url{https://doi.org/10.1063/1.126545}

von Bargen, N. \& Waff, H. S. (1986) Permeabilities, interfacial areas and curvatures of partially molten systems: results of numerical computations of equilibrium microstructures. Journal of Geophysical Research: Solid Earth 91, 9261--9276. \url{https://doi.org/10.1029/JB091iB09p09261}

Waff, H. S. \& Bulau, J. R. (1979) Equilibrium fluid distribution in an ultramafic partial melt under hydrostatic stress conditions. Journal of Geophysical Research: Solid Earth 84, 6109--6114. \url{https://doi.org/10.1029/JB084iB11p06109}

Warren, P. H. (1985) The magma ocean concept and lunar evolution. Annual Review of Earth and Planetary Sciences 13, 201--240. \url{https://doi.org/10.1146/annurev.ea.13.050185.001221}

Warren, P. H. \& Wasson, J. T. (1979) The origin of KREEP. Reviews of Geophysics 17, 73--88. \url{https://doi.org/10.1029/RG017i001p00073}

Welsch, B., Faure, F., Famin, V., Baronnet, A. \& Bachèlery, P. (2013) Dendritic crystallization: a single process for all the textures of olivine in basalts? Journal of Petrology 54, 539--574. \url{https://doi.org/10.1093/petrology/egs077}

Wieser, P. E., Shi, S. C., Gleeson, M. L. M., Rangel, B., DeVitre, C. L., Bearden, A. T., Lynn, K. J. \& Caumon, M.-C. (2025) Fluid inclusion constraints on the geometry of the magmatic plumbing system beneath Mauna Loa --- Part 1: lavas and tephras. Bulletin of Volcanology 87, 89. \url{https://doi.org/10.1007/s00445-025-01874-5}

Zhao, W., Wang, B.-W., Gu, L.-X., Tang, X., Liu, Y., Awais, M., Liu, X.-C., Li, X.-H., Hu, R. \& Li, Q.-L. (2023) Nanoscale evidence of zircon--coffinite--xenotime solid solution observed by atom probe tomography. Chemical Geology 638, 121697. \url{https://doi.org/10.1016/j.chemgeo.2023.121697}

Zhou, Q. et al. (2025) Ultra-depleted mantle source of basalts from the South Pole--Aitken basin. Nature 643, 371--375. \url{https://doi.org/10.1038/s41586-025-09131-7}

\section*{Supplementary Text}

\subsection*{Supplementary Text S1: Candidate rare-earth-related mass windows}
Nominal REE positions are severely overlapped in the olivine APT spectrum. The 2+/3+ atomic-ion, monoxide and dioxide positions fall mainly between \textasciitilde{}44 and \textasciitilde{}90 Da, where Si\textsubscript{x}O\textsubscript{y} and Fe\textsubscript{x}O\textsubscript{y} clusters are abundant, whereas the less-interfered singly charged REE region is flat background. A peak at a nominal REE mass is therefore not sufficient for elemental identification, and the detected field-evaporation ion would not by itself reveal structural bonding or charge compensation.

Natural-abundance templates (Meija et al., 2016) were generated for the fifteen lanthanides plus Y and Sc in 1+, 2+, 3+, monoxide and dioxide forms and compared with the 0.005 Da spectrum. This rejects several earlier single-mass assignments: the feature near 76 Da is consistent with SiO\textsubscript{3}\textsuperscript{+}, and the 164 Da feature with Si\textsubscript{3}O\textsubscript{5}\textsuperscript{+}. Crucially, the 83.96 Da peak does not resolve the predicted \textsuperscript{147}SmO\textsuperscript{2+}, \textsuperscript{149}SmO\textsuperscript{2+} and \textsuperscript{152}SmO\textsuperscript{2+} spacing and relative intensities. Alternative co-segregating molecular clusters involving the independently enriched Ca, Al, P, Na or Ti inventories cannot be excluded.

We also screened 112 nominal REE-related windows for spatial association with the R\'eunion boundary (Table S2). The \textasciitilde{}208 million-ion reconstruction was divided into depth slabs; the local boundary in each slab was located from Ca\textsuperscript{+}, and ion positions were expressed as signed distance to that local boundary. Ca\textsuperscript{+} recovers a strong boundary maximum, whereas SiO\textsubscript{3}\textsuperscript{+} and Si\textsubscript{3}O\textsubscript{5}\textsuperscript{+} remain flat (Figure S8). This validates the spatial localization but not the chemical identity of an overlapping candidate window.

The 83.96 Da candidate window has the largest response (core/matrix ratio 1.35), and several nominal Sc-, Y- and Gd-related windows also deviate from the flat controls. Because 112 candidates were screened, the quoted single-window sigma values are descriptive rather than multiple-testing-corrected identification probabilities. Spatial enrichment proves that a carrier in the window is boundary associated; it does not prove that the nominal REE species is that carrier. Accordingly, Figures 3 and S8--S9 and Table S2 use candidate-window terminology, and the manuscript makes no elemental Sm, Sc, Y or Gd claim. Confirmation would require an interference-free isotope envelope or acquisition conditions that move the REE into less-overlapped channels.

\section*{Supplementary Figures}

\begin{figure}[htbp]
\centering
\includegraphics[width=0.95\textwidth]{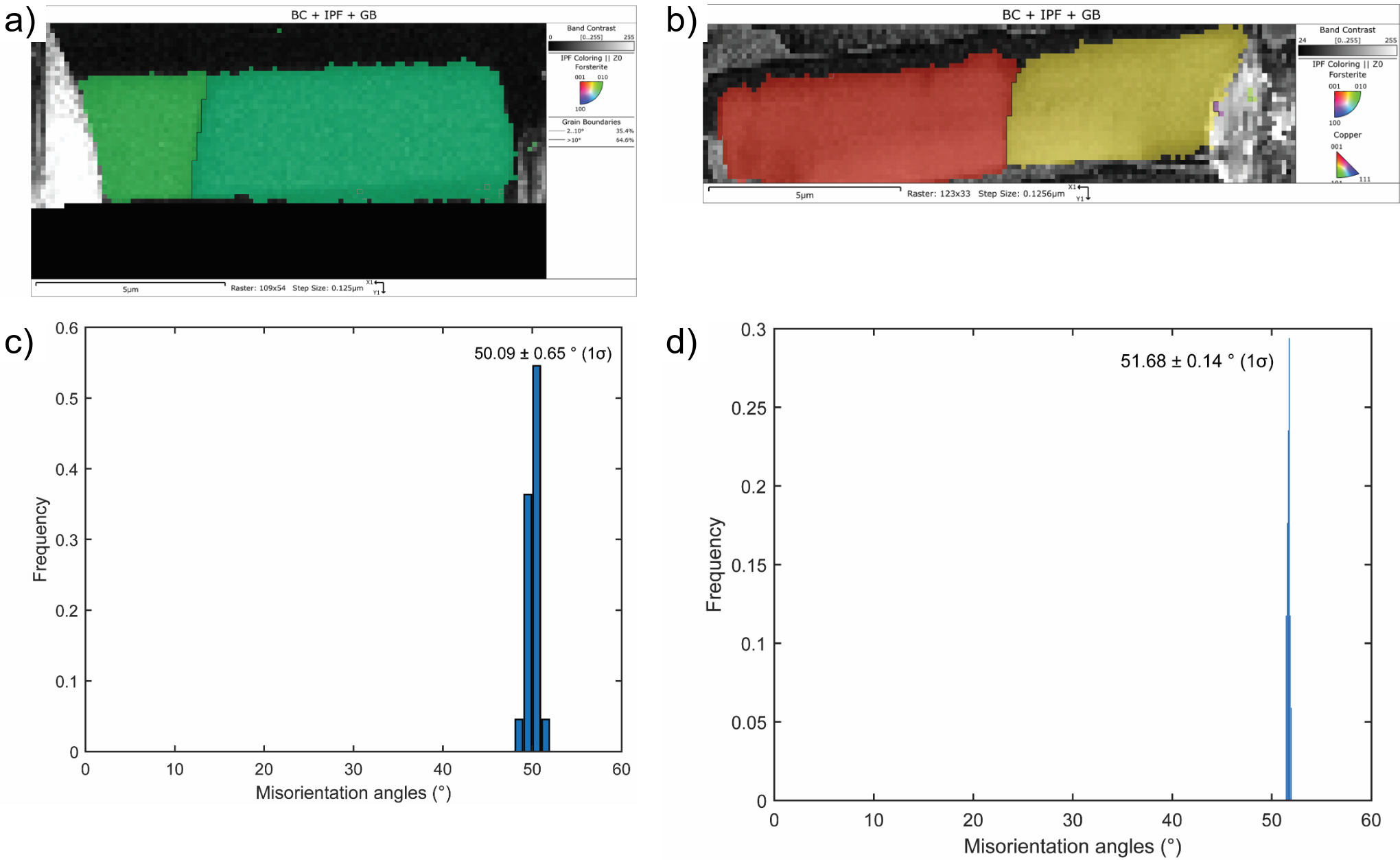}
\caption*{\textbf{Figure S1.} Reflection-EBSD results from the planar surfaces of FIB-prepared lamellae mounted for subsequent TEM: R\'eunion (a, c) and Mauna Loa (b, d). (a, b) Band contrast (BC), inverse-pole-figure (IPF) coloring and grain-boundary (GB) maps acquired with a 0.125 $\mu$m raster step. The raster resolves the orientations of the two grains but does not define the physical width of the boundary. (c, d) Disorientation-angle distributions of mapped boundary segments, reported as mean $\pm1\sigma$: 50.09\textdegree{} $\pm$ 0.65\textdegree{} for R\'eunion and 51.68\textdegree{} $\pm$ 0.14\textdegree{} for Mauna Loa. For Mauna Loa, retained segment operators span 51.52--51.76\textdegree{} about an axis within 2.62--3.17\textdegree{} of $[\bar{1}00]$ and do not match the established olivine twin operators. The R\'eunion archive does not retain rotation-axis/plane metadata, so only a high-angle classification is claimed there.}
\end{figure}
\begin{figure}[htbp]
\centering
\includegraphics[width=\textwidth]{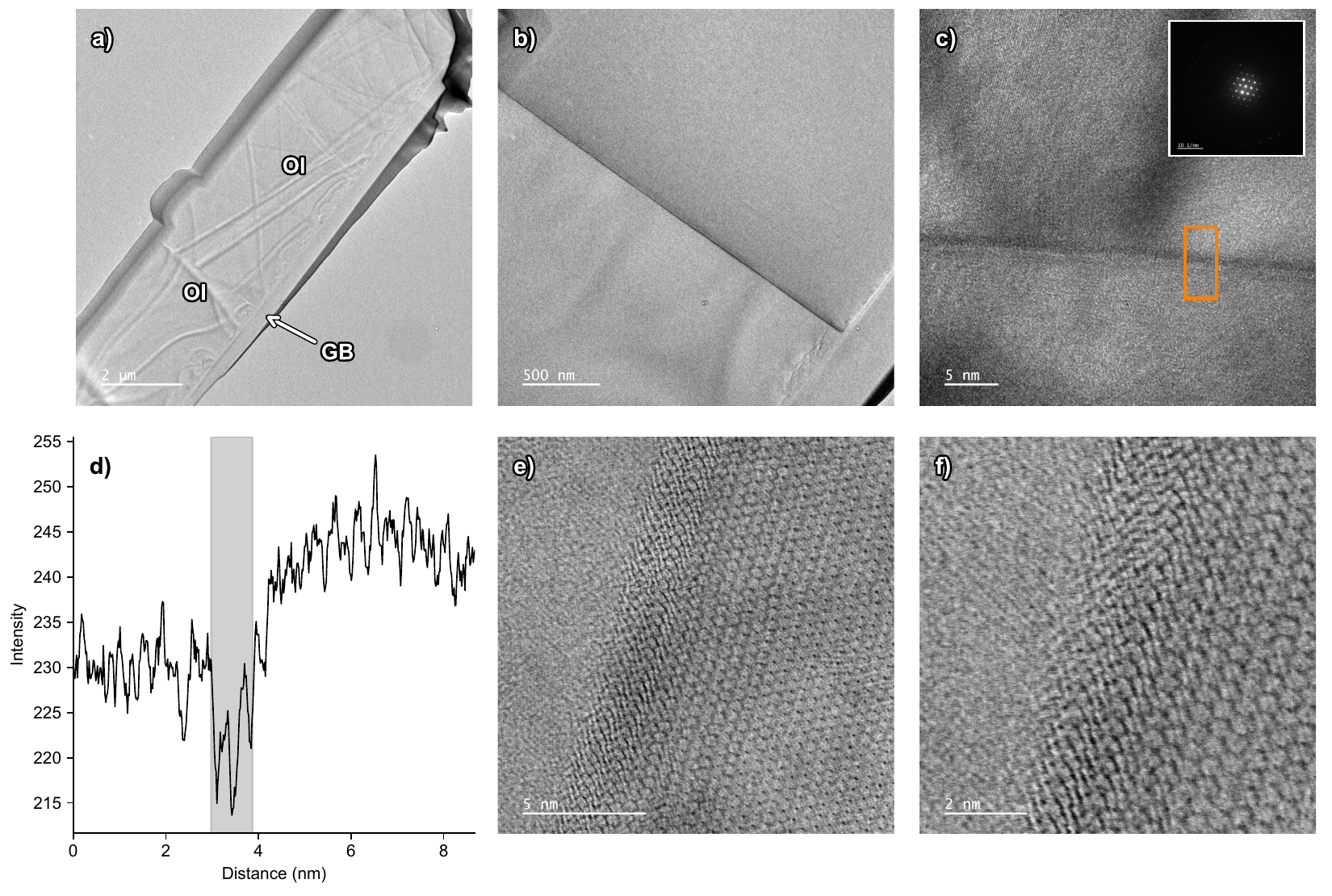}
\caption*{\textbf{Figure S2.} TEM of the Mauna Loa (Hawaii) olivine grain boundary, in the same panel arrangement as Figure 2. (a) Low-magnification bright-field image of the boundary (GB) separating two olivine grains (Ol). (b) Higher-magnification bright-field view. (c) HR-TEM image across the boundary; inset: SAED pattern of the adjacent grain, with the orange box marking the analyzed region. (d) Intensity profile across the boundary (boundary region shaded), giving a structural GB thickness of \textasciitilde{}1.05 nm. (e, f) HR-TEM lattice images of the two adjacent grains.}
\end{figure}
\begin{figure}[htbp]
\centering
\includegraphics[width=0.95\textwidth]{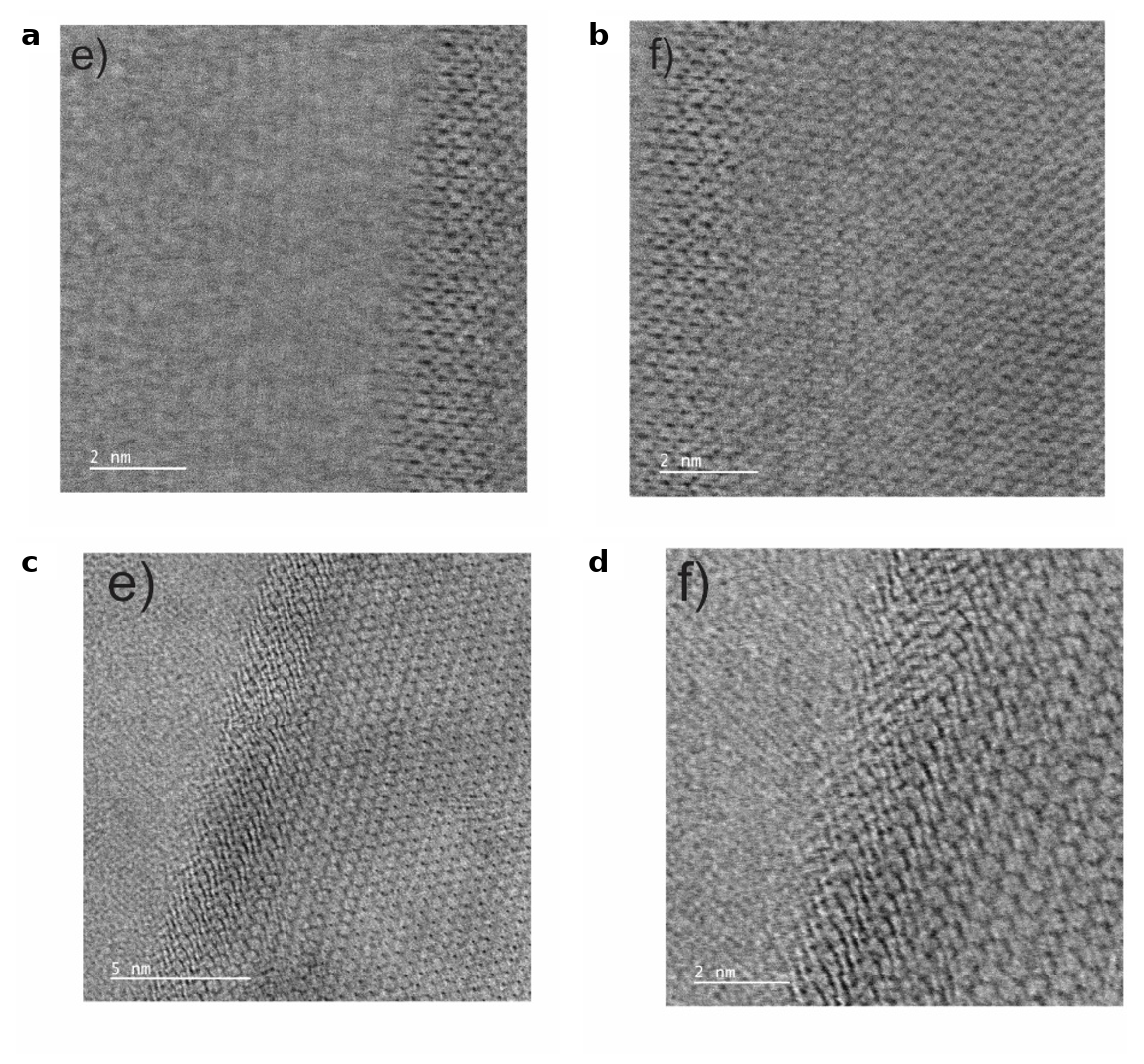}
\caption*{\textbf{Figure S3.} High-resolution STEM images of the olivine grain boundaries: (a, b) R\'eunion and (c, d) Mauna Loa, acquired at different tilt angles for structural observation. Periodic lattice contrast extends to the interface from both grains, and no continuous amorphous or glassy film is resolved within the spatial and projection limits of the images.}
\end{figure}
\begin{figure}[htbp]
\centering
\includegraphics[width=0.95\textwidth]{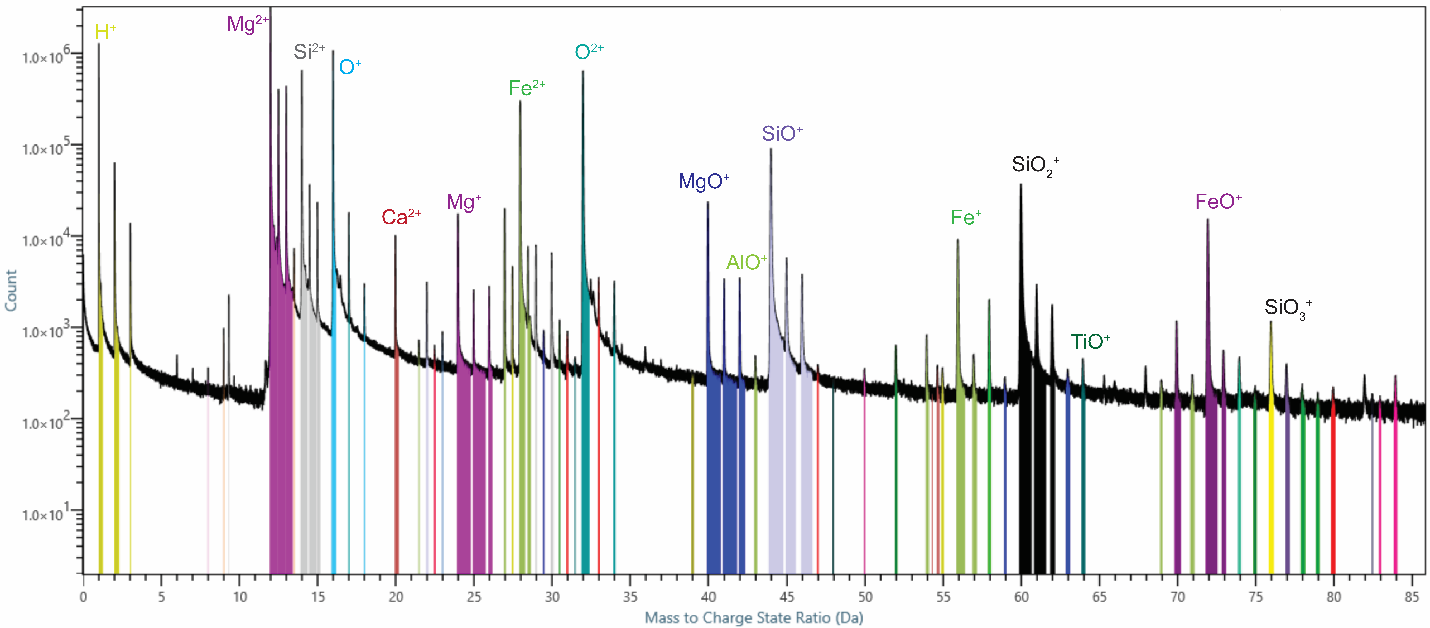}
\caption*{\textbf{Figure S4.} Representative APT mass spectrum of the Réunion olivine, illustrating the mass-to-charge ranging used for ion identification. The spectrum is truncated above 85 Da for clarity, and the principal high-count peaks are annotated with their assigned species; the full ranging is listed in Table S1. The Hawaiian spectrum was indexed with the identical peak ranges.}
\end{figure}
\begin{figure}[htbp]
\centering
\includegraphics[width=0.95\textwidth]{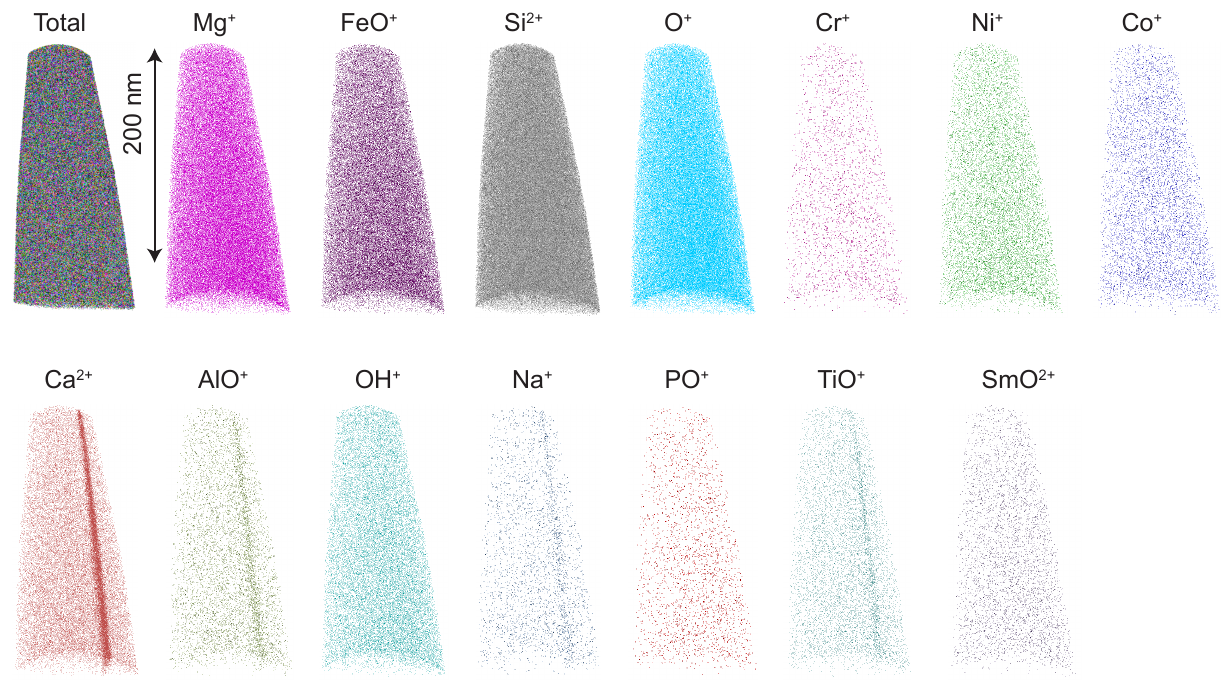}
\caption*{\textbf{Figure S5.} Atom probe tomography (APT) ion maps of the Mauna Loa (Hawaii) olivine needle ($\sim$250 nm long), viewed face-on to the grain boundary so that the boundary plane is seen edge-on. The top-left panel shows all ranged species together (one color per ion type). The remaining top-row maps (Mg\textsuperscript{+}, FeO\textsuperscript{+}, Si\textsuperscript{2+}, O\textsuperscript{+}, Cr\textsuperscript{+}, Ni\textsuperscript{+}, Co\textsuperscript{+}), together with OH\textsuperscript{+}, PO\textsuperscript{+} and the nominal SmO\textsuperscript{2+} candidate window, are homogeneous across the needle, whereas Ca\textsuperscript{2+}, AlO\textsuperscript{+}, Na\textsuperscript{+} and TiO\textsuperscript{+} show pronounced planar enrichment. The nominal SmO\textsuperscript{2+} label is a mass-window assignment, not a confirmed Sm detection.}
\end{figure}
\begin{figure}[htbp]
\centering
\includegraphics[width=0.5\textwidth]{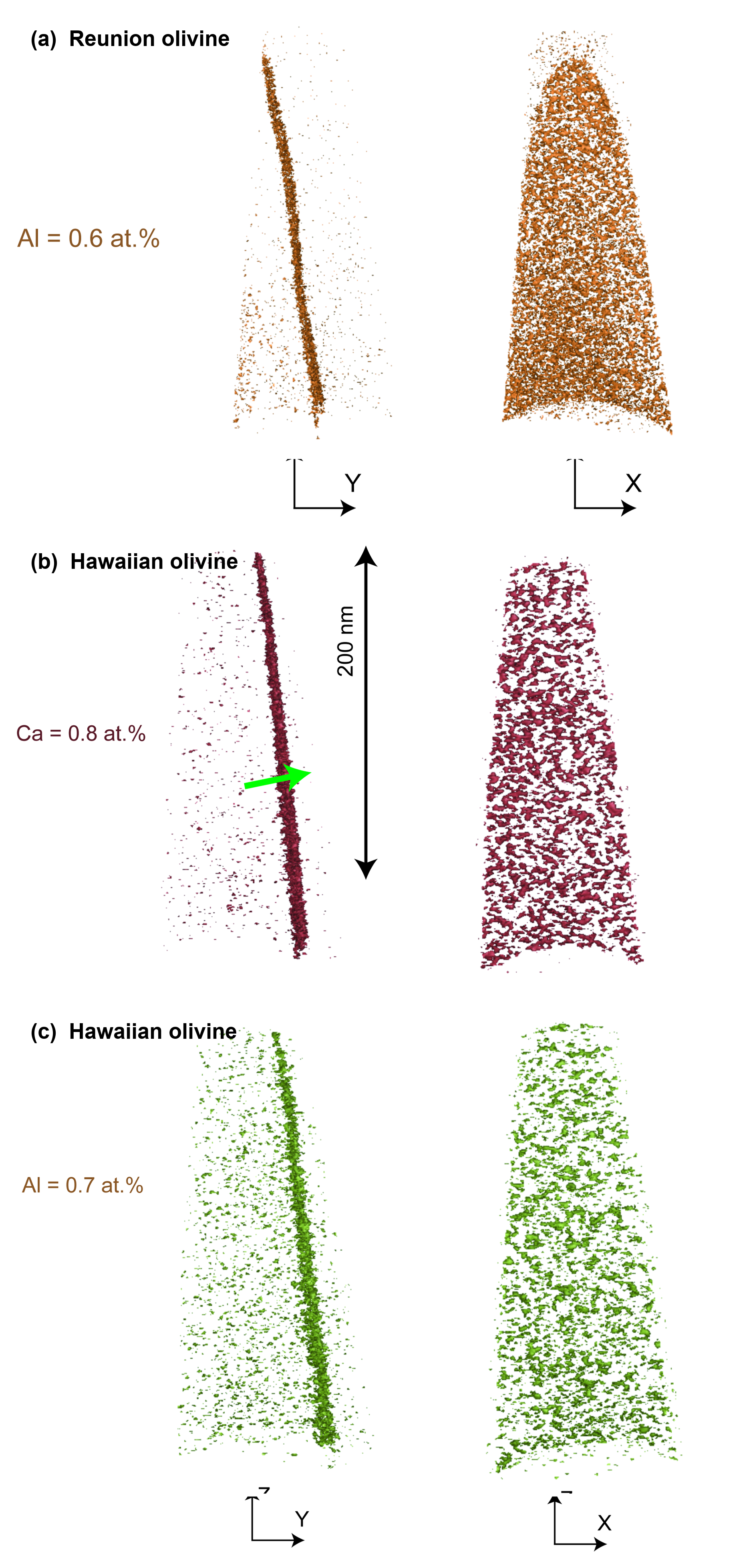}
\caption*{\textbf{Figure S6.} Iso-concentration surface maps obtained by APT, each shown as two orthogonal projections of the reconstruction; axis triads give the spatial orientation. (a) Aluminium (Al) iso-surface at 0.6 at.\% in the Réunion olivine sample, the companion to the Ca surface of Figure 4 (the 1-D profiling direction and 400 nm scale bar are shown there). (b) Calcium (Ca) iso-surface at 0.8 at.\% and (c) aluminium (Al) iso-surface at 0.7 at.\% in the Hawaiian olivine sample, both delineating a well-defined grain boundary; the green marker in (b) indicates the direction along which the 1-D compositional profile (Figure S7) is extracted.}
\end{figure}
\begin{figure}[htbp]
\centering
\includegraphics[width=0.95\textwidth]{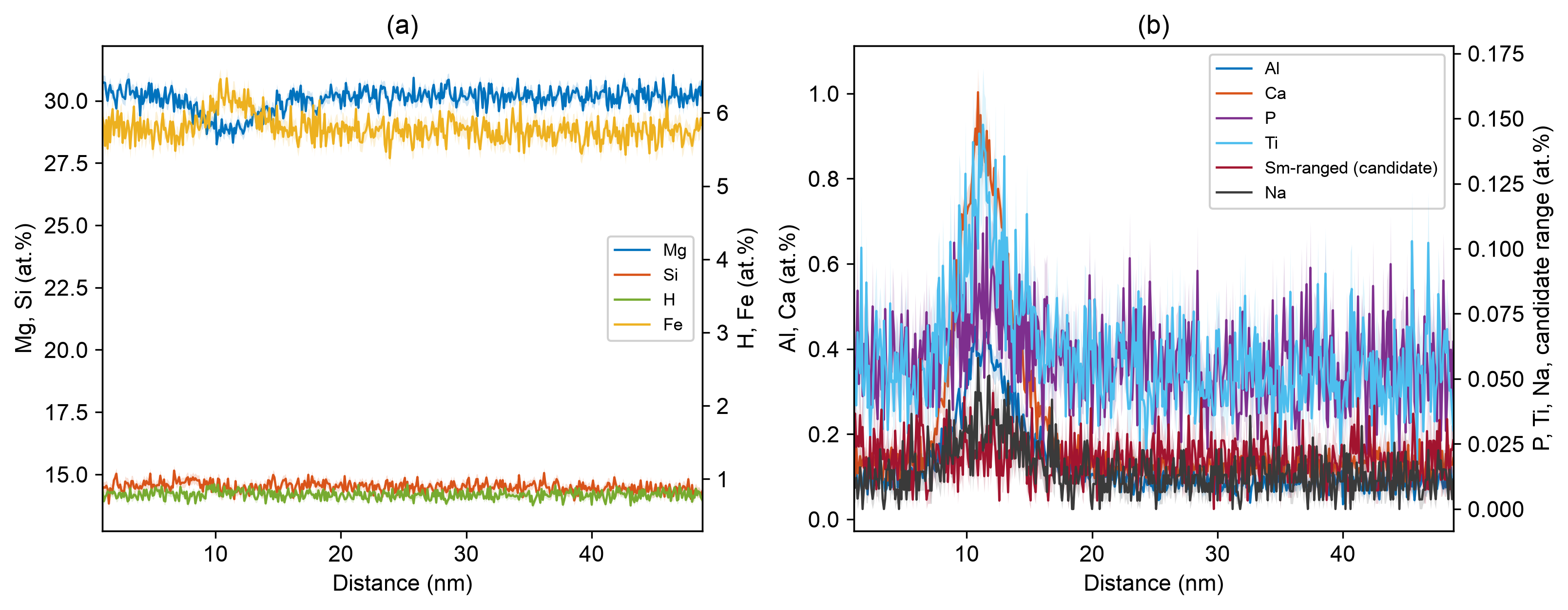}
\caption*{\textbf{Figure S7.} One-dimensional APT profile across the Mauna Loa boundary from R13\_05578-v01, Cube ID 1, X-axis (49.8 nm at 0.10 nm spacing), along the green marker in Figure S6b. Shading denotes reported $\pm1\sigma$ counting statistics. (a) Mg and Si (left axis) with H and total elemental Fe (right axis); Fe increases modestly from 5.75 at.\% in the matrix to a 6.14 at.\% $\pm2$ nm boundary mean. (b) Al and Ca (left axis) with P, Ti, Na and the nominal Sm-ranged candidate signal (right axis). Ca and Al point maxima are 1.00 and 0.45 at.\%, while their $\pm2$ nm means are 0.76 and 0.35 at.\%. The Sm-ranged trace is not a confirmed elemental Sm measurement.}
\end{figure}
\begin{figure}[htbp]
\centering
\includegraphics[width=0.92\textwidth]{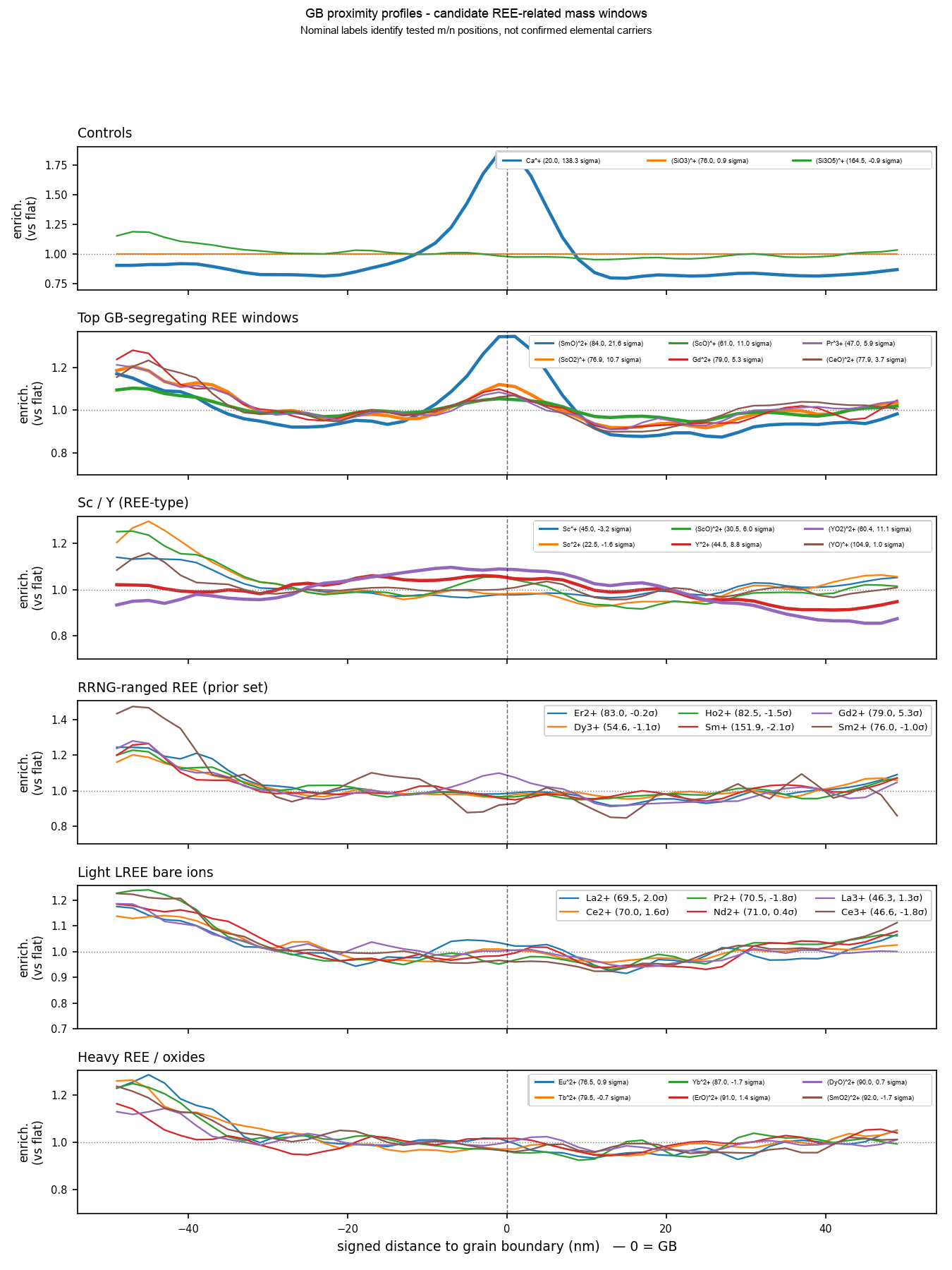}
\caption*{\textbf{Figure S8.} Grain-boundary proximity profiles for candidate REE-related mass windows, grouped by nominal charge/oxide assignment. Each curve is ion density versus signed distance to the locally flattened boundary, normalized to matrix density. A peak at zero establishes boundary association of a carrier within that mass window, not the carrier's elemental identity. Ca\textsuperscript{+} is the segregant control and SiO\textsubscript{3}\textsuperscript{+}/Si\textsubscript{3}O\textsubscript{5}\textsuperscript{+} are flat controls.}
\end{figure}
\begin{landscape}
\begin{figure}[p]
\centering
\includegraphics[width=0.94\linewidth]{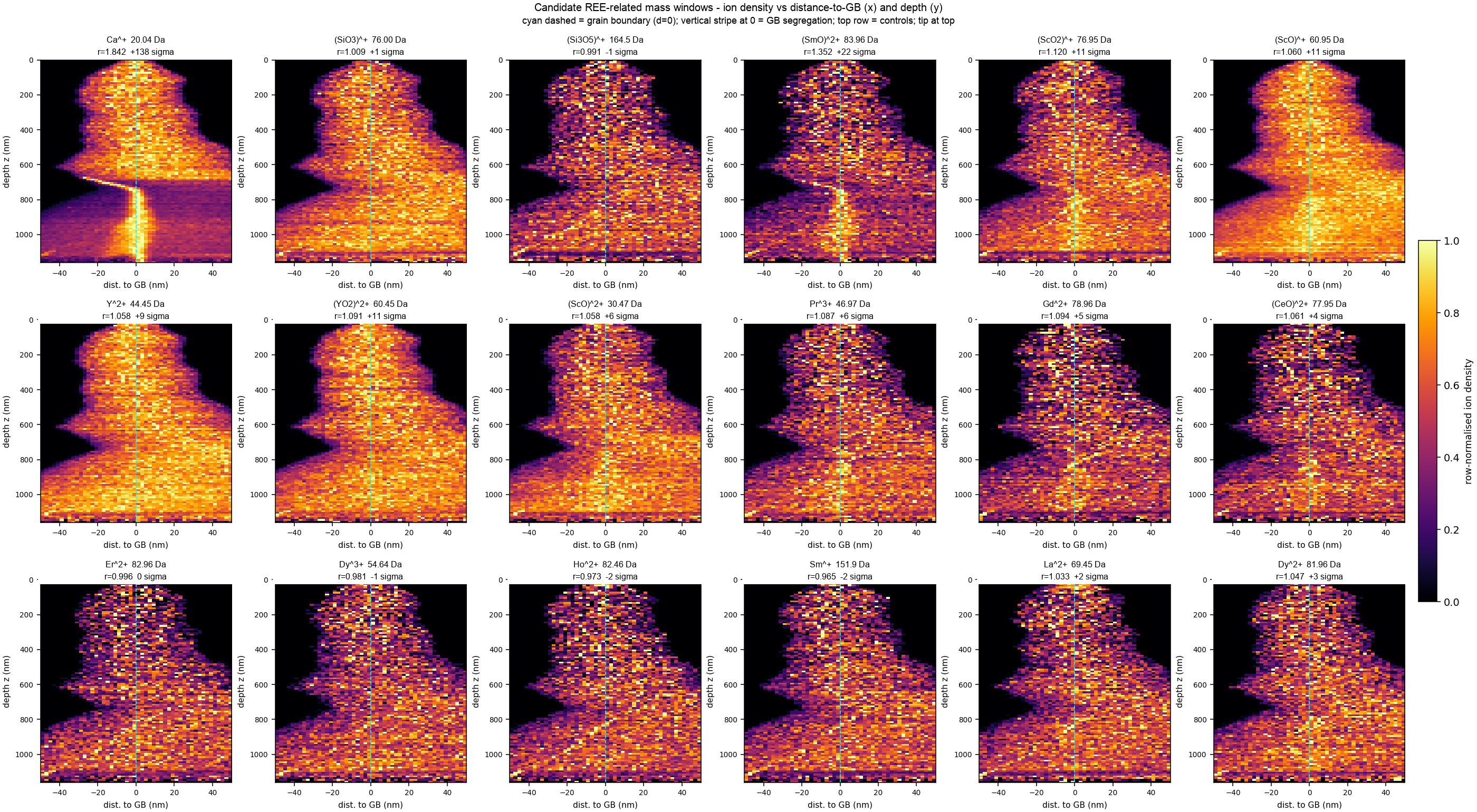}
\caption*{\textbf{Figure S9.} Spatial maps of selected candidate REE-related mass windows, labelled by their nominal assignments, compared with the Ca\textsuperscript{+} segregant and SiO\textsubscript{3}\textsuperscript{+} flat control. Boundary-associated density in a window does not establish that Sm, Sc or Gd is the carrier because diagnostic isotope envelopes are unresolved and molecular-cluster isobars remain possible.}
\end{figure}
\end{landscape}
\begin{figure}[htbp]
\centering
\includegraphics[width=0.95\textwidth]{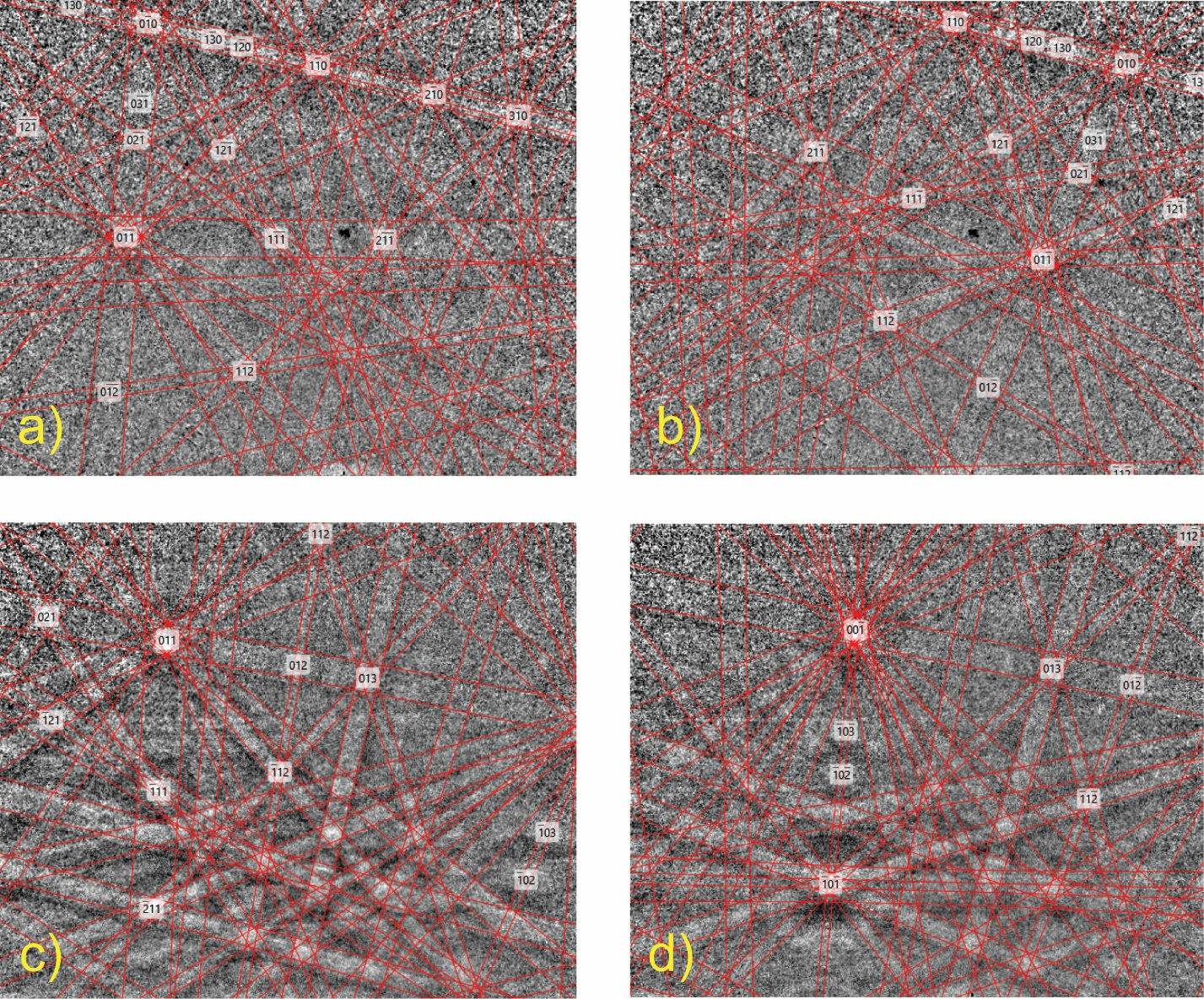}
\caption*{\textbf{Figure S10.} Crystallographic indexing of APT detector-hit-density (desorption) maps generated separately from the two sides of each boundary: (a, b) left and right sides of the Piton de la Fournaise boundary and (c, d) left and right sides of the Mauna Loa boundary. Pole centers and associated zone traces were selected manually and compared with the orthorhombic olivine stereographic geometry. An orientation was retained only when several non-collinear poles and their connecting angular relationships were reproduced simultaneously. Red lines are calculated zone traces for the accepted orientation, and labels give Miller indices; they are not direct images of atomic planes. No spatial-distribution-map or Fourier-transform measurement was used.}
\end{figure}

\section*{Supplementary Tables}

\begin{table}[htbp]
\centering
\caption*{\textbf{Table S1.} Mass-to-charge ($m/c$) ranging used to assign ion and polyion species in the olivine atom-probe spectra, applied identically to the R\'eunion and Hawaiian datasets (cf.\ Figure S4). Charge states are those consistent with each peak's $m/c$ and the species' molecular mass. The matrix molecular clusters (Si\textsubscript{x}O\textsubscript{y}, Fe\textsubscript{x}O\textsubscript{y}) that produce the isobaric interferences discussed in the text are included; rare-earth-related ranges are treated separately in Supplementary Text S1 and are not listed here. Ga originates from Ga\textsuperscript{+} focused-ion-beam preparation.}
\footnotesize
\begin{tabular}{rl@{\hskip 2em}rl}
\hline
$m/c$ (Da) & Ion & $m/c$ (Da) & Ion \\
\hline
1.00--1.26 & H\textsuperscript{1+} & 43.83--44.77 & SiO\textsuperscript{1+} \\
2.00--2.30 & H\textsubscript{2}\textsuperscript{1+} & 44.91--45.57 & SiO\textsuperscript{1+} \\
3.00--3.06 & H\textsubscript{3}\textsuperscript{1+} & 45.90--46.63 & SiO\textsuperscript{1+} \\
7.99--8.01 & Mg\textsuperscript{3+} & 46.92--47.02 & PO\textsuperscript{1+} \\
8.98--9.03 & Al\textsuperscript{3+} & 47.91--48.00 & Ti\textsuperscript{1+} \\
9.31--9.35 & Si\textsuperscript{3+} & 49.91--50.02 & Cr\textsuperscript{1+} \\
11.95--12.44 & Mg\textsuperscript{2+} & 51.91--52.07 & Si\textsubscript{2}O\textsubscript{3}\textsuperscript{2+} \\
12.46--12.94 & Mg\textsuperscript{2+} & 53.88--54.09 & Fe\textsuperscript{1+} \\
12.96--13.40 & Mg\textsuperscript{2+} & 54.87--55.03 & Mn\textsuperscript{1+} \\
13.47--13.52 & Al\textsuperscript{2+} & 55.83--56.40 & Fe\textsuperscript{1+} \\
13.90--14.38 & Si\textsuperscript{2+} & 56.86--57.16 & Fe\textsuperscript{1+} \\
14.45--14.92 & Si\textsuperscript{2+} & 57.85--58.06 & Ni\textsuperscript{1+} \\
14.95--15.18 & Si\textsuperscript{2+} & 58.89--59.03 & Co\textsuperscript{1+} \\
15.89--16.19 & O\textsuperscript{1+} & 59.85--60.67 & SiO\textsubscript{2}\textsuperscript{1+} \\
16.97--17.05 & OH\textsuperscript{1+} & 60.81--61.58 & SiO\textsubscript{2}\textsuperscript{1+} \\
17.98--18.04 & O\textsuperscript{1+} & 61.84--62.16 & SiO\textsubscript{2}\textsuperscript{1+} \\
19.94--20.19 & Ca\textsuperscript{2+} & 62.83--63.12 & PO\textsubscript{2}\textsuperscript{1+} \\
21.47--21.53 & AlO\textsuperscript{2+} & 63.84--64.09 & TiO\textsuperscript{1+} \\
21.95--22.08 & SiO\textsuperscript{2+} & 68.86--69.02 & Ga\textsuperscript{1+} \\
22.96--23.03 & Na\textsuperscript{1+} & 69.79--70.20 & FeO\textsuperscript{1+} \\
23.94--24.80 & Mg\textsuperscript{1+} & 70.80--71.08 & Ga\textsuperscript{1+} \\
24.93--25.75 & Mg\textsuperscript{1+} & 71.78--72.55 & FeO\textsuperscript{1+} \\
25.93--26.18 & Mg\textsuperscript{1+} & 72.81--73.09 & FeO\textsuperscript{1+} \\
26.93--27.05 & Fe\textsuperscript{2+} & 73.85--74.05 & NiO\textsuperscript{1+} \\
27.43--27.52 & Mn\textsuperscript{2+} & 74.84--75.05 & CoO\textsuperscript{1+} \\
27.91--28.30 & Fe\textsuperscript{2+} & 75.83--76.16 & SiO\textsubscript{3}\textsuperscript{1+} \\
28.43--28.64 & Fe\textsuperscript{2+} & 79.82--80.09 & TiO\textsubscript{2}\textsuperscript{1+} \\
28.94--29.03 & Si\textsuperscript{1+} & 87.81--88.34 & Si\textsubscript{2}O\textsubscript{2}\textsuperscript{1+} \\
29.42--29.51 & Co\textsuperscript{2+} & 89.80--90.18 & NiO\textsubscript{2}\textsuperscript{1+} \\
29.92--30.09 & Si\textsuperscript{1+} & 97.83--98.08 & Si\textsubscript{3}O\textsubscript{7}\textsuperscript{2+} \\
30.92--31.03 & P\textsuperscript{1+} & 99.84--100.12 & Si\textsubscript{3}O\textsuperscript{1+} \\
31.44--31.51 & TiO\textsuperscript{2+} & 103.78--104.63 & Si\textsubscript{2}O\textsubscript{3}\textsuperscript{1+} \\
31.89--32.40 & O\textsubscript{2}\textsuperscript{1+} & 104.85--105.68 & Si\textsubscript{2}O\textsubscript{3}\textsuperscript{1+} \\
32.94--33.06 & O\textsubscript{2}H\textsuperscript{1+} & 105.84--106.13 & Si\textsubscript{2}O\textsubscript{3}\textsuperscript{1+} \\
33.92--34.06 & O\textsubscript{2}\textsuperscript{1+} & 119.78--120.62 & Si\textsubscript{2}O\textsubscript{4}\textsuperscript{1+} \\
38.92--39.07 & K\textsuperscript{1+} & 120.81--121.63 & Si\textsubscript{2}O\textsubscript{4}\textsuperscript{1+} \\
39.92--40.76 & MgO\textsuperscript{1+} & 121.84--122.35 & Si\textsubscript{2}O\textsubscript{4}\textsuperscript{1+} \\
40.89--41.79 & MgO\textsuperscript{1+} & 135.82--136.30 & Si\textsubscript{2}O\textsubscript{5}\textsuperscript{1+} \\
41.92--42.32 & MgO\textsuperscript{1+} & 163.57--165.39 & Si\textsubscript{3}O\textsubscript{5}\textsuperscript{1+} \\
42.91--43.10 & AlO\textsuperscript{1+} & 179.75--180.50 & Si\textsubscript{3}O\textsubscript{6}\textsuperscript{1+} \\
\hline
\end{tabular}
\end{table}

\begin{landscape}
\begin{spacing}{1.0}
\scriptsize
\setlength{\tabcolsep}{4pt}
\begin{longtable}{@{}l l l r r r c r r r r r p{2.3cm}@{}}
\caption*{\textbf{Table S2.} Spatial segregation screen of 112 candidate rare-earth-related mass-to-charge windows in the R\'eunion reconstruction (Supplementary Text S1). Element and species fields are nominal mass assignments, not confirmed identifications. For each window the table gives the center and bounds, nominal charge and oxygen count, ion counts within $\pm$50 nm and the $\pm$4 nm core, core/matrix ratio, descriptive single-window $\sigma$, and spatial verdict. The $\sigma$ values are not multiple-testing-corrected; enrichment identifies a boundary-associated carrier within a window but not its chemical species. Ca, SiO$_3$ and Si$_3$O$_5$ are positive/flat controls. The same data are provided as Table\_S2.csv.}\\[2pt]
\hline
El. & Form & Species & $m/n$ & Win$_{lo}$ & Win$_{hi}$ & $q$ & $n_{O}$ & $N_{\pm50}$ & $N_{core}$ & Ratio & $\sigma$ & Verdict \\
\hline
\endfirsthead
\multicolumn{13}{@{}l}{\footnotesize\itshape Table S2 (continued)}\\
\hline
El. & Form & Species & $m/n$ & Win$_{lo}$ & Win$_{hi}$ & $q$ & $n_{O}$ & $N_{\pm50}$ & $N_{core}$ & Ratio & $\sigma$ & Verdict \\
\hline
\endhead
\hline
\endfoot
Sc & M+ (1+) & Sc+ & 44.9554 & 44.6554 & 45.2554 & 1 & 0 & 457059 & 47562 & 0.985 & -3.2 & flat / matrix-like \\
Sc & M2+ (2+) & Sc2+ & 22.4774 & 22.3274 & 22.6274 & 2 & 0 & 86470 & 8978 & 0.983 & -1.6 & flat / matrix-like \\
Sc & M3+ (3+) & Sc3+ & 14.9848 & 14.8848 & 15.0848 & 3 & 0 & 446210 & 45010 & 0.955 & -9.8 & GB-depleted \\
Sc & MO+ (1+) & ScO+ & 60.9503 & 60.6503 & 61.2503 & 1 & 1 & 319804 & 35800 & 1.06 & 11.0 & weak GB enrichment * \\
Sc & MO2+ (2+) & ScO2+ & 30.4749 & 30.3249 & 30.6249 & 2 & 1 & 101147 & 11298 & 1.058 & 6.0 & weak GB enrichment * \\
Sc & MO2 (1+) & ScO2\_+ & 76.9452 & 76.6452 & 77.2452 & 1 & 2 & 75326 & 8908 & 1.12 & 10.7 & GB-segregating candidate window \\
Sc & MO2 (2+) & ScO2\_2+ & 38.4723 & 38.3223 & 38.6223 & 2 & 2 & 62390 & 6399 & 0.971 & -2.3 & flat / matrix-like \\
Y & M+ (1+) & Y+ & 88.9053 & 88.6053 & 89.2053 & 1 & 0 & 81069 & 8546 & 0.998 & -0.2 & flat / matrix-like \\
Y & M2+ (2+) & Y2+ & 44.4524 & 44.3024 & 44.6024 & 2 & 0 & 217798 & 24345 & 1.058 & 8.8 & weak GB enrichment * \\
Y & M3+ (3+) & Y3+ & 29.6347 & 29.5347 & 29.7347 & 3 & 0 & 59508 & 6046 & 0.962 & -3.0 & flat / matrix-like \\
Y & MO+ (1+) & YO+ & 104.9002 & 104.6002 & 105.2002 & 1 & 1 & 78676 & 8401 & 1.011 & 1.0 & flat / matrix-like \\
Y & MO2+ (2+) & YO2+ & 52.4498 & 52.2998 & 52.5998 & 2 & 1 & 37676 & 3933 & 0.988 & -0.7 & flat / matrix-like \\
Y & MO2 (1+) & YO2\_+ & 120.8951 & 120.5951 & 121.1951 & 1 & 2 & 54589 & 6043 & 1.048 & 3.7 & flat / matrix-like \\
Y & MO2 (2+) & YO2\_2+ & 60.4473 & 60.2973 & 60.5973 & 2 & 2 & 140986 & 16241 & 1.091 & 11.1 & GB-segregating (suspicious) ** \\
La & M+ (1+) & La+ & 138.9058 & 138.6058 & 139.2058 & 1 & 0 & 31235 & 3175 & 0.962 & -2.2 & flat / matrix-like \\
La & M2+ (2+) & La2+ & 69.4526 & 69.3026 & 69.6026 & 2 & 0 & 34372 & 3750 & 1.033 & 2.0 & flat / matrix-like \\
La & M3+ (3+) & La3+ & 46.3016 & 46.2016 & 46.4016 & 3 & 0 & 48338 & 5196 & 1.018 & 1.3 & flat / matrix-like \\
La & MO+ (1+) & LaO+ & 154.9007 & 154.6007 & 155.2007 & 1 & 1 & 28833 & 2912 & 0.956 & -2.4 & flat / matrix-like \\
La & MO2+ (2+) & LaO2+ & 77.4501 & 77.3001 & 77.6001 & 2 & 1 & 29057 & 3020 & 0.984 & -0.9 & flat / matrix-like \\
La & MO2 (1+) & LaO2\_+ & 170.8956 & 170.5956 & 171.1956 & 1 & 2 & 26725 & 2734 & 0.969 & -1.7 & flat / matrix-like \\
La & MO2 (2+) & LaO2\_2+ & 85.4475 & 85.2975 & 85.5975 & 2 & 2 & 25052 & 2630 & 0.994 & -0.3 & flat / matrix-like \\
Ce & M+ (1+) & Ce+ & 139.9049 & 139.6049 & 140.2049 & 1 & 0 & 31514 & 3241 & 0.974 & -1.5 & flat / matrix-like \\
Ce & M2+ (2+) & Ce2+ & 69.9522 & 69.8022 & 70.1022 & 2 & 0 & 84582 & 9086 & 1.017 & 1.6 & flat / matrix-like \\
Ce & M3+ (3+) & Ce3+ & 46.6346 & 46.5346 & 46.7346 & 3 & 0 & 43500 & 4471 & 0.973 & -1.8 & flat / matrix-like \\
Ce & MO+ (1+) & CeO+ & 155.8998 & 155.5998 & 156.1998 & 1 & 1 & 29220 & 3085 & 1.0 & -0.0 & flat / matrix-like \\
Ce & MO2+ (2+) & CeO2+ & 77.9496 & 77.7996 & 78.0996 & 2 & 1 & 33709 & 3778 & 1.061 & 3.7 & weak GB enrichment * \\
Ce & MO2 (1+) & CeO2\_+ & 171.8947 & 171.5947 & 172.1947 & 1 & 2 & 26624 & 2730 & 0.971 & -1.5 & flat / matrix-like \\
Ce & MO2 (2+) & CeO2\_2+ & 85.9471 & 85.7971 & 86.0971 & 2 & 2 & 25464 & 2706 & 1.006 & 0.3 & flat / matrix-like \\
Pr & M+ (1+) & Pr+ & 140.9071 & 140.6071 & 141.2071 & 1 & 0 & 31121 & 3195 & 0.972 & -1.6 & flat / matrix-like \\
Pr & M2+ (2+) & Pr2+ & 70.4533 & 70.3033 & 70.6033 & 2 & 0 & 33294 & 3407 & 0.969 & -1.8 & flat / matrix-like \\
Pr & M3+ (3+) & Pr3+ & 46.9687 & 46.8687 & 47.0687 & 3 & 0 & 43318 & 4972 & 1.087 & 5.9 & weak GB enrichment * \\
Pr & MO+ (1+) & PrO+ & 156.902 & 156.602 & 157.202 & 1 & 1 & 28703 & 2967 & 0.979 & -1.2 & flat / matrix-like \\
Pr & MO2+ (2+) & PrO2+ & 78.4507 & 78.3007 & 78.6007 & 2 & 1 & 28031 & 2935 & 0.991 & -0.5 & flat / matrix-like \\
Pr & MO2 (1+) & PrO2\_+ & 172.8969 & 172.5969 & 173.1969 & 1 & 2 & 26191 & 2700 & 0.976 & -1.3 & flat / matrix-like \\
Pr & MO2 (2+) & PrO2\_2+ & 86.4482 & 86.2982 & 86.5982 & 2 & 2 & 24550 & 2606 & 1.005 & 0.3 & flat / matrix-like \\
Nd & M+ (1+) & Nd+ & 141.9072 & 141.6072 & 142.2072 & 1 & 0 & 31412 & 3246 & 0.978 & -1.2 & flat / matrix-like \\
Nd & M2+ (2+) & Nd2+ & 70.9533 & 70.8033 & 71.1033 & 2 & 0 & 39232 & 4167 & 1.006 & 0.4 & flat / matrix-like \\
Nd & M3+ (3+) & Nd3+ & 47.302 & 47.202 & 47.402 & 3 & 0 & 37248 & 3827 & 0.973 & -1.7 & flat / matrix-like \\
Nd & MO+ (1+) & NdO+ & 157.9021 & 157.6021 & 158.2021 & 1 & 1 & 949 & 95 &  &  & low-stats (untestable) \\
Nd & MO2+ (2+) & NdO2+ & 78.9508 & 78.8008 & 79.1008 & 2 & 1 & 909 & 93 &  &  & low-stats (untestable) \\
Nd & MO2 (1+) & NdO2\_+ & 173.897 & 173.597 & 174.197 & 1 & 2 & 851 & 90 &  &  & low-stats (untestable) \\
Nd & MO2 (2+) & NdO2\_2+ & 86.9482 & 86.7982 & 87.0982 & 2 & 2 & 765 & 83 &  &  & low-stats (untestable) \\
Sm & M+ (1+) & Sm+ & 151.9192 & 151.6192 & 152.2192 & 1 & 0 & 34094 & 3476 & 0.965 & -2.1 & flat / matrix-like \\
Sm & M2+ (2+) & Sm2+ & 75.9593 & 75.8093 & 76.1093 & 2 & 0 & 2163 & 214 & 0.937 & -1.0 & flat / matrix-like \\
Sm & M3+ (3+) & Sm3+ & 50.6394 & 50.5394 & 50.7394 & 3 & 0 & 30177 & 3144 & 0.986 & -0.8 & flat / matrix-like \\
Sm & MO (1+) & (SmO)\textsuperscript{+} & 167.9141 & 167.6141 & 168.2141 & 1 & 1 & 27793 & 2931 & 0.999 & -0.1 & flat / matrix-like \\
Sm & MO (2+) & (SmO)\textsuperscript{2+} & 83.9568 & 83.8068 & 84.1068 & 2 & 1 & 35712 & 5098 & 1.352 & 21.6 & GB-segregating candidate window \\
Sm & MO2 (1+) & (SmO\textsubscript{2})\textsuperscript{+} & 183.909 & 183.609 & 184.209 & 1 & 2 & 25314 & 2572 & 0.962 & -2.0 & flat / matrix-like \\
Sm & MO2 (2+) & (SmO\textsubscript{2})\textsuperscript{2+} & 91.9542 & 91.8042 & 92.1042 & 2 & 2 & 34502 & 3544 & 0.973 & -1.7 & flat / matrix-like \\
Eu & M+ (1+) & Eu+ & 152.9207 & 152.6207 & 153.2207 & 1 & 0 & 31234 & 3087 & 0.936 & -3.7 & flat / matrix-like \\
Eu & M2+ (2+) & Eu2+ & 76.4601 & 76.3101 & 76.6101 & 2 & 0 & 31940 & 3424 & 1.015 & 0.9 & flat / matrix-like \\
Eu & M3+ (3+) & Eu3+ & 50.9732 & 50.8732 & 51.0732 & 3 & 0 & 29990 & 3235 & 1.021 & 1.2 & flat / matrix-like \\
Eu & MO+ (1+) & EuO+ & 168.9156 & 168.6156 & 169.2156 & 1 & 1 & 692 & 69 &  &  & low-stats (untestable) \\
Eu & MO2+ (2+) & EuO2+ & 84.4575 & 84.3075 & 84.6075 & 2 & 1 & 829 & 76 &  &  & low-stats (untestable) \\
Eu & MO2 (1+) & EuO2\_+ & 184.9105 & 184.6105 & 185.2105 & 1 & 2 & 612 & 58 &  &  & low-stats (untestable) \\
Eu & MO2 (2+) & EuO2\_2+ & 92.455 & 92.305 & 92.605 & 2 & 2 & 593 & 59 &  &  & low-stats (untestable) \\
Gd & M+ (1+) & Gd+ & 157.9236 & 157.6236 & 158.2236 & 1 & 0 & 28651 & 2938 & 0.971 & -1.6 & flat / matrix-like \\
Gd & M2+ (2+) & Gd2+ & 78.9615 & 78.8115 & 79.1115 & 2 & 0 & 30257 & 3496 & 1.094 & 5.3 & weak GB enrichment * \\
Gd & M3+ (3+) & Gd3+ & 52.6408 & 52.5408 & 52.7408 & 3 & 0 & 28931 & 2973 & 0.973 & -1.5 & flat / matrix-like \\
Gd & MO+ (1+) & GdO+ & 173.9185 & 173.6185 & 174.2185 & 1 & 1 & 852 & 65 &  &  & low-stats (untestable) \\
Gd & MO2+ (2+) & GdO2+ & 86.959 & 86.809 & 87.109 & 2 & 1 & 792 & 65 &  &  & low-stats (untestable) \\
Gd & MO2 (1+) & GdO2\_+ & 189.9134 & 189.6134 & 190.2134 & 1 & 2 & 840 & 88 &  &  & low-stats (untestable) \\
Gd & MO2 (2+) & GdO2\_2+ & 94.9564 & 94.8064 & 95.1064 & 2 & 2 & 815 & 86 &  &  & low-stats (untestable) \\
Tb & M+ (1+) & Tb+ & 158.9248 & 158.6248 & 159.2248 & 1 & 0 & 27793 & 2914 & 0.993 & -0.4 & flat / matrix-like \\
Tb & M2+ (2+) & Tb2+ & 79.4621 & 79.3121 & 79.6121 & 2 & 0 & 27342 & 2850 & 0.987 & -0.7 & flat / matrix-like \\
Tb & M3+ (3+) & Tb3+ & 52.9746 & 52.8746 & 53.0746 & 3 & 0 & 29230 & 2994 & 0.97 & -1.7 & flat / matrix-like \\
Tb & MO+ (1+) & TbO+ & 174.9197 & 174.6197 & 175.2197 & 1 & 1 & 1101 & 116 &  &  & low-stats (untestable) \\
Tb & MO2+ (2+) & TbO2+ & 87.4596 & 87.3096 & 87.6096 & 2 & 1 & 782 & 87 &  &  & low-stats (untestable) \\
Tb & MO2 (1+) & TbO2\_+ & 190.9146 & 190.6146 & 191.2146 & 1 & 2 & 989 & 101 &  &  & low-stats (untestable) \\
Tb & MO2 (2+) & TbO2\_2+ & 95.457 & 95.307 & 95.607 & 2 & 2 & 798 & 94 &  &  & low-stats (untestable) \\
Dy & M+ (1+) & Dy+ & 163.9286 & 163.6286 & 164.2286 & 1 & 0 & 0 & 0 &  &  & low-stats (untestable) \\
Dy & M2+ (2+) & Dy2+ & 81.964 & 81.814 & 82.114 & 2 & 0 & 37943 & 4196 & 1.047 & 3.0 & flat / matrix-like \\
Dy & M3+ (3+) & Dy3+ & 54.6425 & 54.5425 & 54.7425 & 3 & 0 & 33253 & 3445 & 0.981 & -1.1 & flat / matrix-like \\
Dy & MO+ (1+) & DyO+ & 179.9235 & 179.6235 & 180.2235 & 1 & 1 & 33123 & 3505 & 1.002 & 0.1 & flat / matrix-like \\
Dy & MO2+ (2+) & DyO2+ & 89.9615 & 89.8115 & 90.1115 & 2 & 1 & 51722 & 5513 & 1.009 & 0.7 & flat / matrix-like \\
Dy & MO2 (1+) & DyO2\_+ & 195.9185 & 195.6185 & 196.2185 & 1 & 2 & 23938 & 2391 & 0.946 & -2.7 & flat / matrix-like \\
Dy & MO2 (2+) & DyO2\_2+ & 97.959 & 97.809 & 98.109 & 2 & 2 & 26252 & 2801 & 1.01 & 0.5 & flat / matrix-like \\
Ho & M+ (1+) & Ho+ & 164.9298 & 164.6298 & 165.2298 & 1 & 0 & 0 & 0 &  &  & low-stats (untestable) \\
Ho & M2+ (2+) & Ho2+ & 82.4646 & 82.3146 & 82.6146 & 2 & 0 & 29251 & 3006 & 0.973 & -1.5 & flat / matrix-like \\
Ho & M3+ (3+) & Ho3+ & 54.9762 & 54.8762 & 55.0762 & 3 & 0 & 32206 & 3419 & 1.005 & 0.3 & flat / matrix-like \\
Ho & MO+ (1+) & HoO+ & 180.9247 & 180.6247 & 181.2247 & 1 & 1 & 28022 & 2931 & 0.99 & -0.5 & flat / matrix-like \\
Ho & MO2+ (2+) & HoO2+ & 90.4621 & 90.3121 & 90.6121 & 2 & 1 & 36857 & 3938 & 1.012 & 0.7 & flat / matrix-like \\
Ho & MO2 (1+) & HoO2\_+ & 196.9196 & 196.6196 & 197.2196 & 1 & 2 & 23466 & 2481 & 1.001 & 0.1 & flat / matrix-like \\
Ho & MO2 (2+) & HoO2\_2+ & 98.4595 & 98.3095 & 98.6095 & 2 & 2 & 22835 & 2347 & 0.973 & -1.3 & flat / matrix-like \\
Er & M+ (1+) & Er+ & 165.9298 & 165.6298 & 166.2298 & 1 & 0 & 30626 & 3299 & 1.02 & 1.1 & flat / matrix-like \\
Er & M2+ (2+) & Er2+ & 82.9646 & 82.8146 & 83.1146 & 2 & 0 & 28031 & 2948 & 0.996 & -0.2 & flat / matrix-like \\
Er & M3+ (3+) & Er3+ & 55.3096 & 55.2096 & 55.4096 & 3 & 0 & 26443 & 2769 & 0.991 & -0.5 & flat / matrix-like \\
Er & MO+ (1+) & ErO+ & 181.9247 & 181.6247 & 182.2247 & 1 & 1 & 26850 & 2807 & 0.99 & -0.5 & flat / matrix-like \\
Er & MO2+ (2+) & ErO2+ & 90.9621 & 90.8121 & 91.1121 & 2 & 1 & 36585 & 3950 & 1.022 & 1.4 & flat / matrix-like \\
Er & MO2 (1+) & ErO2\_+ & 197.9196 & 197.6196 & 198.2196 & 1 & 2 & 23250 & 2437 & 0.992 & -0.4 & flat / matrix-like \\
Er & MO2 (2+) & ErO2\_2+ & 98.9595 & 98.8095 & 99.1095 & 2 & 2 & 22528 & 2341 & 0.984 & -0.8 & flat / matrix-like \\
Tm & M+ (1+) & Tm+ & 168.9337 & 168.6337 & 169.2337 & 1 & 0 & 26951 & 2697 & 0.947 & -2.8 & flat / matrix-like \\
Tm & M2+ (2+) & Tm2+ & 84.4666 & 84.3166 & 84.6166 & 2 & 0 & 25806 & 2659 & 0.976 & -1.3 & flat / matrix-like \\
Tm & M3+ (3+) & Tm3+ & 56.3109 & 56.2109 & 56.4109 & 3 & 0 & 30792 & 3277 & 1.008 & 0.4 & flat / matrix-like \\
Tm & MO+ (1+) & TmO+ & 184.9286 & 184.6286 & 185.2286 & 1 & 1 & 25239 & 2669 & 1.001 & 0.1 & flat / matrix-like \\
Tm & MO2+ (2+) & TmO2+ & 92.464 & 92.314 & 92.614 & 2 & 1 & 33187 & 3255 & 0.929 & -4.2 & GB-depleted \\
Tm & MO2 (1+) & TmO2\_+ & 200.9235 & 200.6235 & 201.2235 & 1 & 2 & 22996 & 2349 & 0.967 & -1.6 & flat / matrix-like \\
Tm & MO2 (2+) & TmO2\_2+ & 100.4615 & 100.3115 & 100.6115 & 2 & 2 & 21772 & 2177 & 0.947 & -2.6 & flat / matrix-like \\
Yb & M+ (1+) & Yb+ & 173.9383 & 173.6383 & 174.2383 & 1 & 0 & 25966 & 2711 & 0.989 & -0.6 & flat / matrix-like \\
Yb & M2+ (2+) & Yb2+ & 86.9689 & 86.8189 & 87.1189 & 2 & 0 & 24545 & 2508 & 0.967 & -1.7 & flat / matrix-like \\
Yb & M3+ (3+) & Yb3+ & 57.9791 & 57.8791 & 58.0791 & 3 & 0 & 100688 & 10130 & 0.953 & -4.9 & GB-depleted \\
Yb & MO+ (1+) & YbO+ & 189.9332 & 189.6332 & 190.2332 & 1 & 1 & 24326 & 2540 & 0.989 & -0.6 & flat / matrix-like \\
Yb & MO2+ (2+) & YbO2+ & 94.9663 & 94.8163 & 95.1163 & 2 & 1 & 24742 & 2574 & 0.985 & -0.8 & flat / matrix-like \\
Yb & MO2 (1+) & YbO2\_+ & 205.9282 & 205.6282 & 206.2282 & 1 & 2 & 22588 & 2246 & 0.941 & -2.9 & flat / matrix-like \\
Yb & MO2 (2+) & YbO2\_2+ & 102.9638 & 102.8138 & 103.1138 & 2 & 2 & 21476 & 2317 & 1.022 & 1.0 & flat / matrix-like \\
Lu & M+ (1+) & Lu+ & 174.9402 & 174.6402 & 175.2402 & 1 & 0 & 25853 & 2652 & 0.971 & -1.5 & flat / matrix-like \\
Lu & M2+ (2+) & Lu2+ & 87.4698 & 87.3198 & 87.6198 & 2 & 0 & 24202 & 2477 & 0.969 & -1.6 & flat / matrix-like \\
Lu & M3+ (3+) & Lu3+ & 58.313 & 58.213 & 58.413 & 3 & 0 & 25630 & 2772 & 1.024 & 1.3 & flat / matrix-like \\
Lu & MO+ (1+) & LuO+ & 190.9351 & 190.6351 & 191.2351 & 1 & 1 & 24364 & 2432 & 0.945 & -2.8 & flat / matrix-like \\
Lu & MO2+ (2+) & LuO2+ & 95.4673 & 95.3173 & 95.6173 & 2 & 1 & 24020 & 2473 & 0.975 & -1.3 & flat / matrix-like \\
Lu & MO2 (1+) & LuO2\_+ & 206.9301 & 206.6301 & 207.2301 & 1 & 2 & 22634 & 2279 & 0.953 & -2.3 & flat / matrix-like \\
Lu & MO2 (2+) & LuO2\_2+ & 103.4648 & 103.3148 & 103.6148 & 2 & 2 & 20800 & 2096 & 0.954 & -2.2 & flat / matrix-like \\
Ca & CONTROL & Ca\_ctrl & 20.04 & 19.94 & 20.193 & 1 & 0 & 255881 & 49767 & 1.842 & 138.3 & segregant control \\
SiO3 & CONTROL & SiO3\_ctrl & 76.0 & 75.826 & 76.158 & 1 & 3 & 105692 & 11259 & 1.009 & 0.9 & flat control \\
Si3O5 & CONTROL & Si3O5\_ctrl & 164.5 & 163.575 & 165.39 & 1 & 5 & 107178 & 11222 & 0.991 & -0.9 & flat control \\
\hline
\end{longtable}
\end{spacing}
\end{landscape}

\end{document}